\documentclass[useAMS]{mn2e} \voffset=-1cm \usepackage{times}
\usepackage{psfig} \usepackage{epsfig} \usepackage{amssymb}
\usepackage{amsbsy}

\def\etal{{\it et al.\ }} 
 \def\kmps{${\rm\thinspace
km}{\rm\thinspace s}^{-1}$}   \def\omegam{{\Omega_{\rm m}}}
\def\beqn{\vspace{2mm} \begin{eqnarray}} \def\eeqn{\vspace{2mm}
\end{eqnarray}} \def\beq{\vspace{2mm} \begin{equation}}
\def\eeq{\vspace{2mm} \end{equation}} \def\Lsun{\hbox{$\rm\thinspace
L_{\odot}$}}

\begin{document}

\title{The Dipole Anisotropy of the 2 Micron All-Sky Redshift Survey}

\author[Erdo\u{g}du \etal] {
\parbox[t]{\textwidth}{P. Erdo\u{g}du$^{1,2}$, J.P. Huchra$^{3}$,
O. Lahav$^{2,4}$, M. Colless$^{5}$, R.M. Cutri$^{6}$ E. Falco$^{3}$,
T. George$^{7}$, T. Jarrett$^{6}$, D. H. Jones$^{8}$,
C.S. Kochanek$^{9}$, L. Macri$^{10}$, J. Mader$^{11}$,
N. Martimbeau$^{3}$, M. Pahre$^{3}$, Q. Parker$^{12}$,
A. Rassat$^{4}$, W. Saunders$^{5,13}$ }
\vspace*{6pt} \\ $^{1}$Department of Physics, Middle East Technical
University, 06531, Ankara, Turkey \\ 
$^{2}$School of Physics \& Astronomy, University of Nottingham,
University Park, Nottingham, NG7 2RD, UK\\
$^{3}$Harvard-Smithsonian Centre of Astrophysics, 60 Garden
Street, MS-20, Cambridge, MA 02138, USA\\ $^{4}$Department of Physics
and Astronomy, University College London, Gower Street, London WC1E
6BT, UK\\ $^{5}$Anglo-Australian Observatory, PO Box 296, Epping,NSW
2052, Australia\\ $^{6}$Infrared Processing and Analysis Center,
California Institute of Technology, Pasadena, CA 91125, USA\\
$^{7}$California Institute of Technology, 4800 Oak Grove Drive,
302-231, Pasadena, CA 91109, USA\\ $^{8}$Research School of Astronomy
and Astrophysics, Mount Stromlo, and Siding Spring Observatories,
Cotter Road, Weston Creek ACT 2611, Australia\\ $^{9}$Department of
Astronomy, Ohio State University, 4055 McPherson Lab, 140 West 18th
Avenue, Columbus, OH 43221, USA\\ $^{10}$National Optical Astronomy
Observatory, 950 North Cherry Avenue, Tucson, AZ 85726, USA\\
$^{11}$W.M. Keck Observatory, Kamuela, HI 96743, USA\\
$^{12}$Department of Physics, Macquarie University, Sydney, NWS 2109,
Australia\\ $^{13}$Royal Observatory, Blackford Hill, Edinburgh, EH9
3HJ, UK } \maketitle

\begin{abstract}
We estimate the acceleration on the Local Group (LG) from the Two
Micron All Sky Redshift Survey (2MRS). 
The sample used includes about 23,200 galaxies with extinction corrected 
magnitudes brighter than $K_{\rm s}=11.25$ and it 
allows us to calculate the flux weighted dipole. The near-infrared f\mbox{}l\mbox{}ux weighted dipoles are very robust because they closely approximate a mass weighted dipole, bypassing 
the effects of redshift distortions and require no preferred reference frame. 
This is combined with the redshift information to determine the
change in dipole with distance.  The misalignment angle between the LG
and the CMB dipole drops to 12$^\circ\pm$7$^\circ$ at around 50
$h^{-1} {\rm Mpc}$, but then increases at larger distances, reaching 
21$^\circ\pm$8$^\circ$ at around 130 $h^{-1} {\rm Mpc}$. 
Exclusion of the galaxies Maffei 1, Maffei 2, Dwingeloo 1, IC342 and M87 brings the resultant flux dipole to 14$^\circ\pm$7$^\circ$ away from 
the CMB velocity dipole
In both cases, the dipole seemingly converges by 60 $h^{-1} {\rm Mpc}$. 
Assuming convergence, the comparison of the 2MRS f\mbox{}l\mbox{}ux dipole and the CMB dipole 
provides a value for  
the combination of the mass density  
and luminosity bias parameters  $\omegam^{0.6}/b_{\rm L}= 0.40 \pm 0.09$. 
\end{abstract}

\section{Introduction}

The most popular mechanism for the formation of large-scale structure
and motions in the Universe is the gravitational growth of primordial
density perturbations.  According to this paradigm, if the density
perturbations are small enough to be approximated by a linear theory,
then the peculiar acceleration vector ${\bf g}({\bf r})$ induced by
the matter distribution around position ${\bf r}$ is related to the
mass by
\begin{equation}
{\bf g}({\bf r})= G\bar{\rho} \int\limits_{\bf r}^{\infty} d^3{\bf
r}^{\prime} \delta_m({{\bf r}^{\prime}}) \frac{{\bf r}^{\prime}-{\bf
r}}{|{\bf r}^{\prime}-{\bf r}|^3}
\label{eqn:g(r)}
\end{equation}
where $\bar{\rho}$ is the mean matter density and $\delta_m({\bf r})$
= $(\rho_m({\bf r})-\bar{\rho})/\bar{\rho}$ is the density contrast of
the mass perturbations.  In linear theory, the peculiar velocity
field, ${\bf v}({\bf r})$, is proportional to the peculiar acceleration:
\begin{equation}
{\bf v}({\bf r})= \frac{H_0 f(\Omega_{\rm m})}{4 \pi G \bar{\rho}}
{\bf g}({\bf r})=\frac{2f(\Omega_{\rm m})}{3 H_0\Omega_{\rm m}}{\bf
g}({\bf r}),
\label{eqn:v(r)}
\end{equation}
where $H_0$ = 100 $h$ ${\rm km} {\rm s}^{-1} {\rm Mpc}^{-1}$ is the
Hubble constant and $f(\Omega_{\rm m})\approx\Omega_{\rm m}^{0.6}$ is
the logarithmic derivative of the amplitude of the growing mode of the
perturbations in mass with respect to the scale factor (Peebles 1980). 
The factor $f(\Omega_{\rm m})$ is only 
weakly dependent on the cosmological constant (Lahav \etal 1991).

During the past twenty five years, particular attention has been paid to the
study of the gravitational acceleration and the peculiar velocity
vectors of the Local Group (LG) of galaxies. It is now widely accepted
that the cosmic microwave background (CMB) dipole is a Doppler effect
arising from the motion of the Sun (but see e.g. Gunn 1988 and
Paczy\'{n}ski \& Piran 1990 who argue that the CMB dipole is of
primordial origin). In this case, the dipole anisotropy of the CMB is
a direct and accurate measurement of the LG peculiar velocity
(c.f. Conklin 1969 and Henry 1971).  The LG acceleration can also be
estimated using surveys of the galaxies tracing the density
inhomogeneities responsible for the acceleration. By comparing the CMB
velocity vector with the acceleration vector\footnote{Both the CMB
velocity and the gravitational acceleration on the LG have units of
velocity and are commonly referred to as `dipoles'. Hereafter, the
terms `LG velocity' and `LG dipole' will be used
interchangeably.}
obtained from the
galaxy surveys, it is possible to investigate the cause of the LG
motion and its cosmological implications. This technique was first 
applied by Yahil, Sandage \& Tammann (1980) using the Revised Shapley-Ames
catalogue and later by Davis \& Huchra (1982) using the CfA
catalogue. Both catalogues were two-dimensional and the analyses were
done using galaxy f\mbox{}l\mbox{}uxes. Since both the gravity and the f\mbox{}l\mbox{}ux are
inversely proportional to the square of the distance, the dipole
vector can be calculated by summing the f\mbox{}l\mbox{}ux vectors and assuming an
average value for the mass-to-light ratio. Lahav (1987) applied the
same method to calculate the dipole anisotropy using maps based on
three galaxy catalogues, UGC, ESO, MCG. The most recent application of the 
galaxy flux dipole analysis was carried out by Maller \etal (2003),
using the two-dimensional Two Micron All-Sky Survey (2MASS) extended
source catalogue (XSC), with a limiting magnitude of $K_{\rm s}=13.57$. 
They found that the LG dipole direction is 16$^\circ$ away that of 
the CMB.
  
Our ability to study the LG motion was greatly advanced by the
whole-sky galaxy samples derived from $IRAS$ {\it Galaxy Catalogues}.
Yahil, Walker \& Rowan-Robinson (1986), Meiksin \& Davis (1986),
Harmon, Lahav \& Meurs (1987), Villumsen \& Strauss (1987) and Lahav,
Rowan-Robinson \& Lynden-Bell (1988) used the two-dimensional $IRAS$
data to obtain the LG dipole.  The dipole vectors
derived by these authors are in agreement with each other and the CMB
dipole vector to within 10$^\circ$-30$^\circ$ degrees.  The inclusion
of galaxy redshifts in the dipole analyses allowed the estimation of
the distance at which most of the peculiar velocity of the LG is
generated ({\it the convergence depth}). However, the estimates of the
convergence depth from various data sets have not agreed. 
Strauss \etal (1992,  $IRAS$ sample), 
Webster, Lahav \& Fisher (1997, $IRAS$ sample), Lynden-Bell, Lahav \&
Burstein (1989, optical sample) and da Costa \etal (2000, a sample of early-type galaxies) suggested that the LG acceleration is mostly
due to galaxies $\lesssim 50 h^{-1} {\rm Mpc}$, while other authors such as
Scaramella, Vettolani \& Zamorani
(1994, Abell/ACO cluster sample), Branchini \& Plionis (1996, Abell/ACO cluster sample), Kocevski \etal (2004) and Kocevski \& Ebeling 
(2005, both using samples of X-ray clusters) 
claimed that there is significant contribution to the dipole from
depths of up to $\approx$200 $h^{-1} {\rm Mpc}$.

Dipole analyses are often used to estimate the 
combination of matter density  and biasing parameters
$\omegam$ and $b$.  In theory, one can equate the velocity
inferred from the CMB measurements with the value derived from a
galaxy survey and obtain a value for $\beta$.  In practice, however,
the galaxy surveys do not measure the true total velocity due their
finite depth (e.g. Lahav, Kaiser \& Hoffman 1990 and Juszkiewicz,
Vittorio \& Wyse 1990).  The true ${\bf v}_{LG}$, as obtained from the
CMB dipole, arises from structure on all scales including structures
further away than the distance a galaxy survey can accurately
measure. Furthermore, the magnitude/f\mbox{}l\mbox{}ux/diameter limit of the survey
and any completeness variations over the sky introduce selection
effects and biases to the calculations. These effects amplify the
errors at large distances (for redshift surveys) and faint magnitudes
(for two dimensional surveys) where the sampling of the galaxy
distribution becomes more sparse. This, combined with the fact that we
sample discretely from an underlying continuous mass distribution
leads to an increase in shot noise error. There may also be a
significant contribution to the dipole from galaxies behind the
Galactic Plane ({\it the zone of avoidance}).  The analysis of the
convergence of the dipole is further complicated by the redshift
distortions on small and large scales which introduce systematic
errors to the derived dipole ({\it the rocket effect}, Kaiser
1987). The following sections discuss these effects in the context of
two different models of biasing.

In this paper, we use the Two Micron All-Sky Redshift Survey (2MRS, Huchra \etal 2005)
\footnote{This work is based on
observations made at the Cerro Tololo Interamerican Observatory (CTIO),
operated for the US National Science Foundation by the Association of
Universities for Research in Astronomy.} to study the LG dipole. The inclusion of the redshift data allows the
calculation of the selection effects of the survey as a function of
distance and enables the study of convergence and thus improves the
analysis of Maller \etal (2003). The paper is structured as follows: The
Two Micron Redshift Survey is described in Section~\ref{sec:dip:data}.
Section 3 discusses the method used in the analysis including the
different weighting schemes, the rocket effect and the choice of
reference frames.  The results are presented in
Section~\ref{sec:dip:results}. The final section includes some
concluding remarks and plans for future work. 

\section{The Two Micron All-Sky Redshift Survey}\label{sec:dip:data}

\begin{figure*}
\psfig{figure=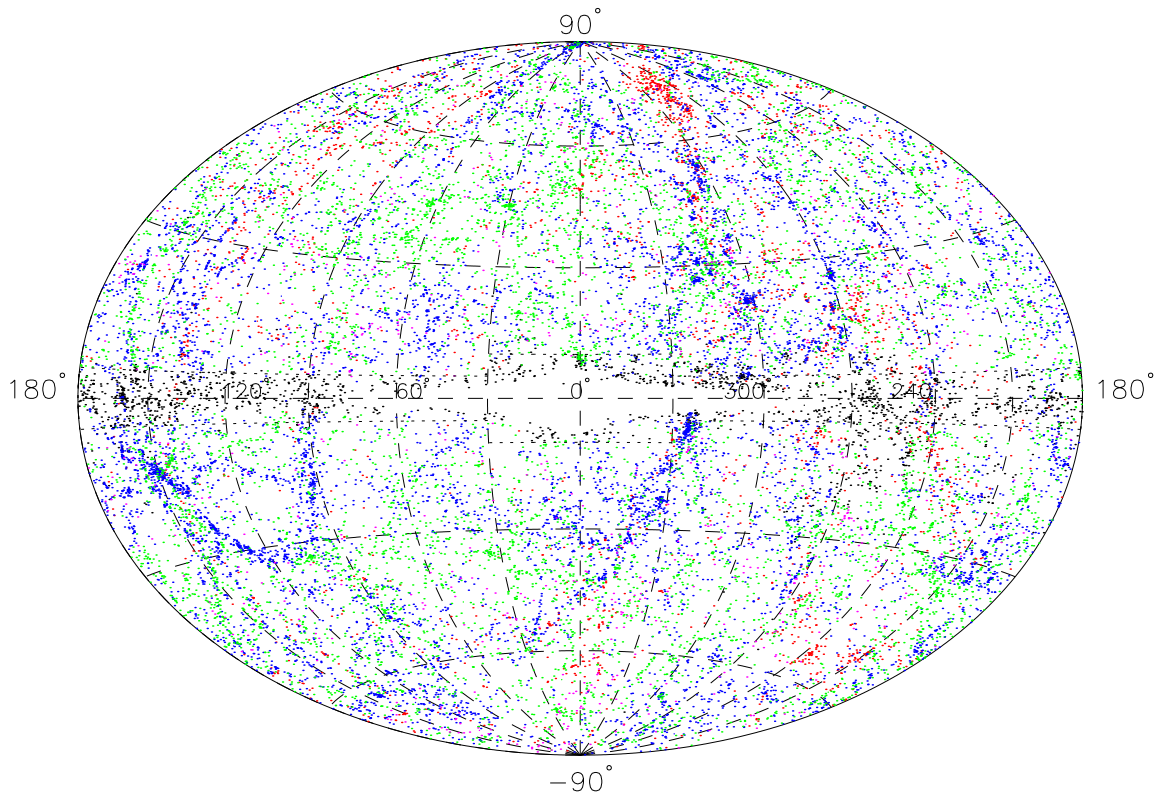,angle=0,width=\textwidth,clip=}
\caption[All Objects in the 2MASS Redshift Catalogue in Galactic
Aitoff Projection] {All Objects in the 2MASS Redshift Catalogue in
Galactic Aitoff Projection. Galaxies with ${\rm z}\leq0.01$ are
plotted in red, $0.01<{\rm z}\le0.025$ are plotted in blue,
$0.025<{\rm z}<0.05$ are plotted in green and ${\rm z}\geq0.05$ are
plotted in magenta. Galaxies without measured redshifts are plotted in
black. The masked region is outlined by dashed lines.}
\label{fig:2MRS}
\end{figure*}

The Two Micron All-Sky Redshift Survey (2MRS) is the densest all-sky
redshift survey to date. The
galaxies in the northern celestial hemisphere are being observed
mainly by the FLWO 1.5-m telescope and at low latitudes by the CTIO. 
In the southern hemisphere, most 
galaxies are observed as a part of the six degree field galaxy survey 
(6dFGS, Jones \etal 2004) conducted by the Anglo Australian Observatory. 
The first phase of the 2MRS is now
completed. In this phase we obtained redshifts for approximately
23,000 2MASS galaxies from a total sample of about 24,800 galaxies with
extinction corrected magnitudes (Schlegel, Finkbeiner \& Davis 1998)
brighter than $K_{\rm s}=11.25$. This magnitude limit corresponds to a
median redshift of $z\approx0.02$ ($\approx 60 h^{-1}$ Mpc).  The
majority of the 1600 galaxies that remain without redshifts are at
very low galactic latitudes or obscured/confused by the dust and the
high stellar density towards the Galactic
Centre. Figure~\ref{fig:2MRS} shows all the objects in the 2MRS in
Galactic Aitoff Projection. Galaxies with ${\rm z}\leq0.01$ are
plotted in red, $0.01<{\rm z}\le0.025$ are plotted in blue,
$0.025<{\rm z}<0.05$ are plotted in green and ${\rm z}\geq0.05$ are
plotted in magenta. Galaxies without measured redshifts are plotted in
black. The 2MRS can be compared with the deeper 2MASS galaxy catalogue
(K$<$14th mag) shown in Jarrett (2004, Figure 1).
\subsection{Survey Completeness}\label{sec:2masscomp}
The 2MASS\footnote{The 2MASS
database and the full documentation are available on the WWW at
http//www.ipac.caltech.edu/2mass.} 
has great photometric uniformity and an unprecedented
integral sky coverage. The photometric uniformity is better than
$4\%$ over the sky including the celestial poles (e.g. Jarrett \etal 2000$a$,
2003). The uniform completeness of the galaxy sample is limited by the
presence of the foreground stars. For a typical high latitude sky less
than $2\%$ of the area is masked by stars. These missing regions are
accounted for using a coverage map, defined as the fraction of the
area of an 8\arcmin$\times$8\arcmin pixel that is not obscured by
stars brighter than 10th mag.  Galaxies are then weighted by the
inverse of the completeness although the analysis is almost unaffected
by this process as the completeness ratio is very close to one for
most parts of the sky.

The stellar contamination of the catalogue is low and is reduced
further by manually inspecting the objects below a redshift of
$cz=200$\kmps.  
The foreground stellar confusion is highest at low
Galactic latitudes, resulting in decreasing overall completeness of
the 2MASS catalogue (e.g. Jarrett \etal 2000$b$) and consequently the
2MRS sample\footnote{See Maller \etal (2005) who 
reduce the stellar contamination in the 2MASS XSC 
by cross-correlating stars with galaxy density.}. 
Stellar confusion also produces colour bias in the 2MASS
galaxy photometry (Cambresy, Jarrett \& Beichman 2005) but this
bias should not be significant for the 2MRS because of its relatively bright 
magnitude limit.

\begin{figure*}
$\begin{array}{c}
\psfig{figure=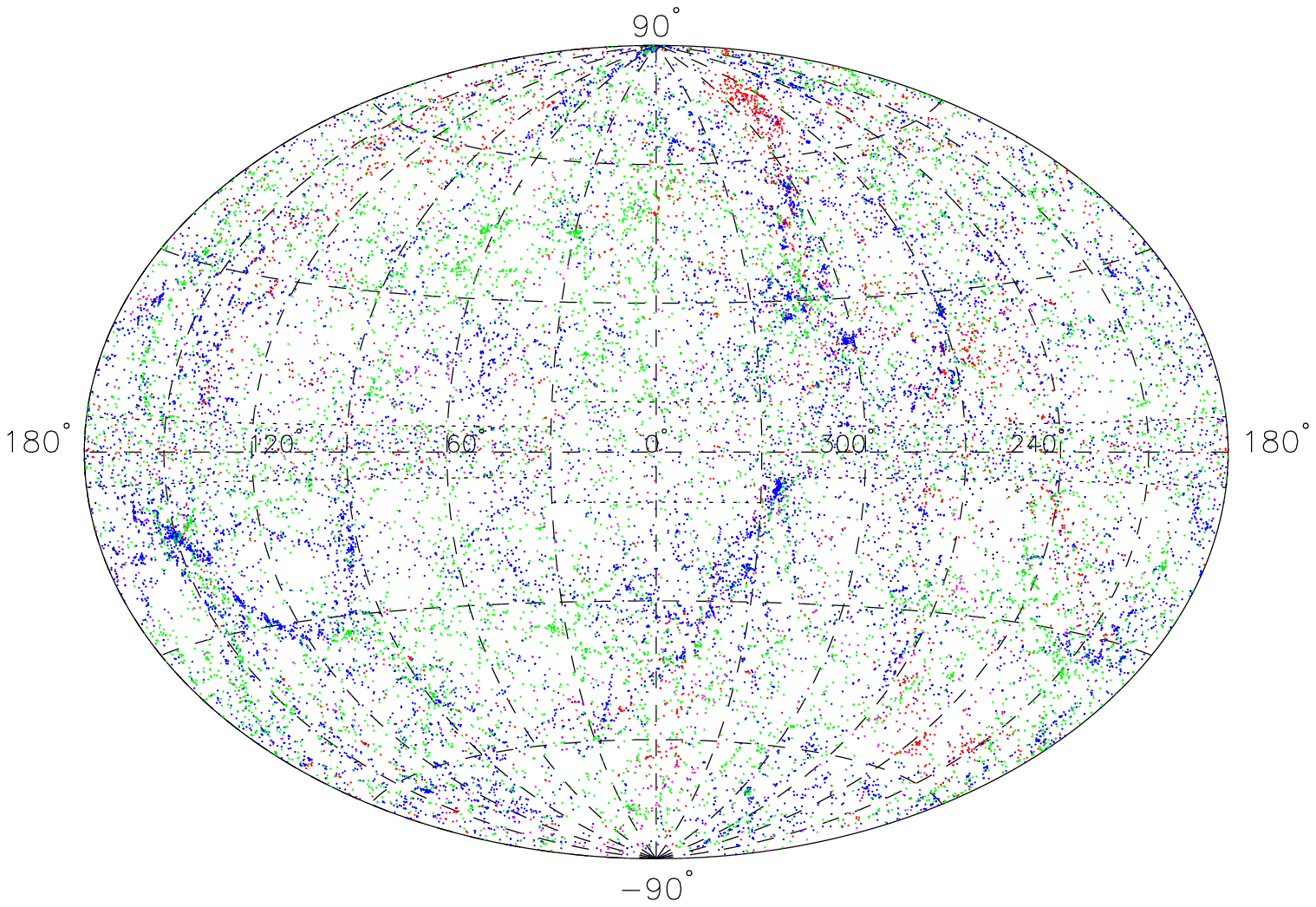,angle=0,width=0.9\textwidth,
height=75mm, clip=} \\
\psfig{figure=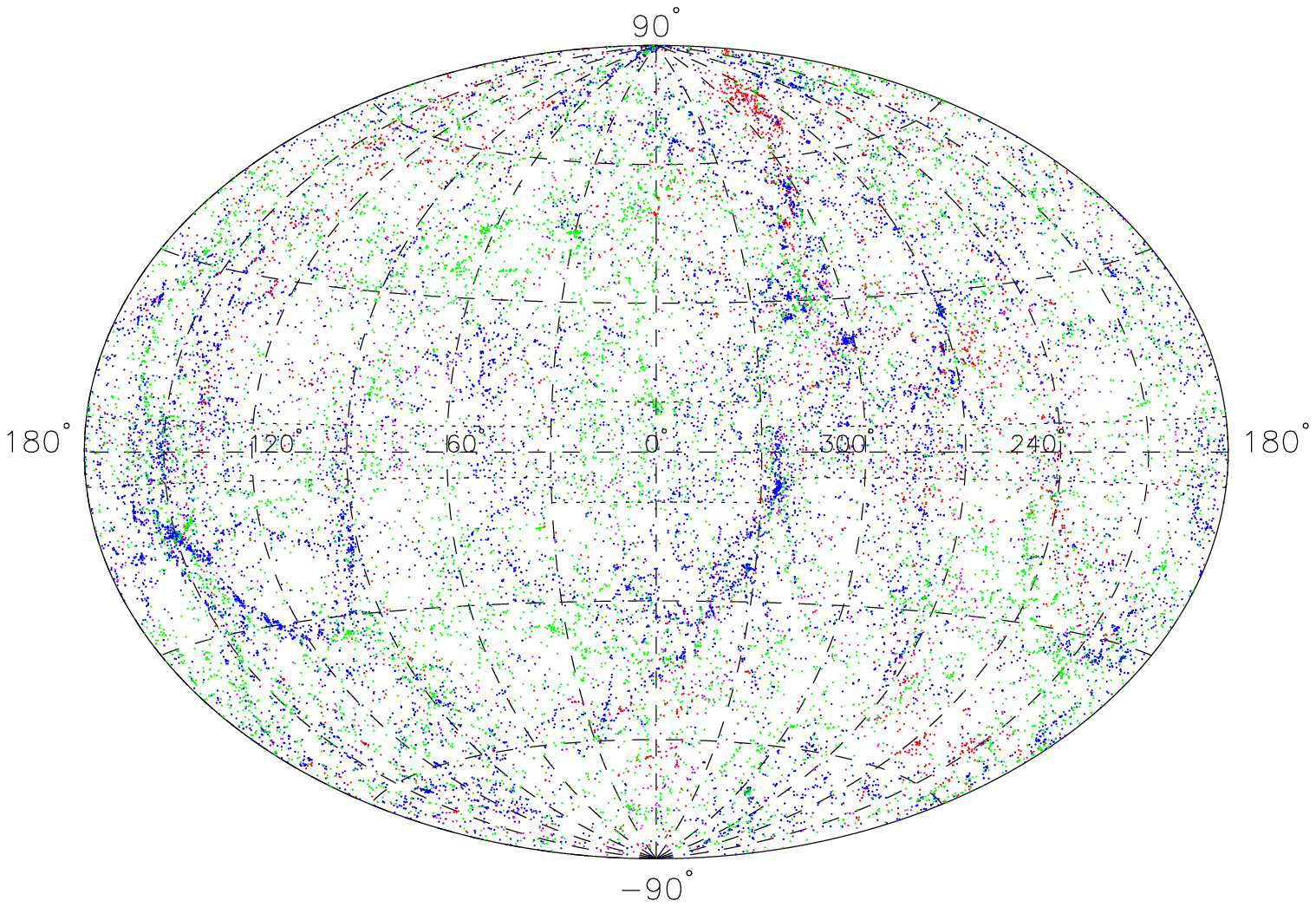,angle=0,width=0.9\textwidth,
height=75mm, clip=} \\
\end{array}$
\caption[Objects in the 2MRS including the random galaxies] {Top:
Objects in the 2MASS Redshift Catalogue used in the analysis in a
Galactic Aitoff Projection, including the random galaxies generated by
using the first technique. Galaxies with ${\rm z}\leq0.01$ are
plotted in red, $0.01<{\rm z}\le0.025$ are plotted in blue,
$0.025<{\rm z}<0.05$ are plotted in green and ${\rm z}\geq0.05$ are
plotted in magenta.Bottom: Same as the top plot but now including the random
galaxies generated in using the second technique. The regions inside
the dashed lines in the plots are masked out and replaced with the
random galaxies shown in the plots. There are 21510 galaxies in each
plot.}
\label{fig:2MRS+mask}
\end{figure*}

In order to account for incompleteness at low Galactic latitudes
we fill the Zone of Avoidance (the plane where 
$|b|<5^\circ$ and $|b|<10^\circ$ in the region $|l|<30^\circ$) with galaxies. 
We keep the galaxies with observed redshifts and apply 
two different methods to compensate for the unobserved (masked) sky:
\begin{itemize}
\item {\bf Method 1}: The masked region is filled with galaxies whose
f\mbox{}l\mbox{}uxes and redshifts are chosen randomly from the whole data
set. These galaxies are placed at random locations within the masked
area. The masked region has the same average density of galaxies as
the rest of the sky.
\item {\bf Method 2}: The masked region is filled following Yahil
\etal (1991). The area is divided into 36 bins of $10^\circ$ in
longitude.  In each angular bin, the distance is divided into bins of 1000 \kmps. The galaxies in each longitude/distance bin are then
sampled from the corresponding longitude/distance bins in the adjacent
strips $-|b_{masked}|-10^\circ<b<|b_{masked}|+10^\circ$ (where
$|b_{masked}|=5^\circ$ or $|b_{masked}|=10^\circ$).  These galaxies
are then placed in random latitudes within the mask region.  This
procedure gives similar results to the more elaborate method of Wiener
reconstruction across the zone of avoidance (Lahav \etal 1994). The
number of galaxies in each masked bin is set to a random Poisson
deviate whose mean equals to the mean number of galaxies in the
adjacent unmasked strips. This procedure is carried out to mimic the
shot noise effects.
\end{itemize}
In reality, the shape of the Zone of Avoidance is not as symmetric as defined in this paper 
with Galactic Bulge centred at $l\approx+5^\circ$ and with latitude offsets 
(see Kraan-Korteweg 2005). However, since we keep 
the galaxies in the masked regions, 
our dipole determinations
should not be greatly influenced by assuming a symmetric mask. 
We test this by changing the centre of the Galactic bulge. 
We confirm that our results are not affected. 
Figure~\ref{fig:2MRS+mask} shows the 2MRS galaxies used in the
analyses in a Galactic Aitoff projection. The galaxies in masked
regions are generated using the first (top plot) and the second method
(bottom plot).

\subsection{Magnitude and F\mbox{}l\mbox{}ux Conversions}

The 2MRS uses the 2MASS magnitude $K_{20}$, 
which is defined\footnote{Column 17
(k$\_$m$\_$k20fc) in the 2MASS XSC}
as the magnitude
inside the circular isophote corresponding to a surface brightness of
$\mu_{K_s}=20 {\rm mag}$ arcsec$^{-2}$ (e.g. Jarrett \etal 2000$a$).
The isophotal magnitudes underestimate the total luminosity by $10\%$
for the early-type and $20\%$ for the late-type galaxies (Jarrett
\etal 2003).  Following Kochanek \etal (2001, Appendix), the offset of
$\Delta=-0.20\pm0.04$ is added to the $K_{20}$ magnitudes.
The galaxy magnitudes are corrected for Galactic extinction using the
dust maps of Schlegel, Finkbeiner \& Davis (1998) and an extinction
correction coefficient of $R_K=0.35$ (Cardelli, Clayton \& Mathis 
1989). As expected, the extinction corrections are small for the 2MRS
sample. The $K_{\rm s}$ band $k$-correction is derived by Kochanek
\etal (2001) based on the stellar population models of Worthey (1994).
The k-correction of $k(z)=-6.0\log(1+z)$, is independent of galaxy
type and valid for $z\lesssim 0.25$.

The f\mbox{}l\mbox{}uxes $S$ are computed from the apparent magnitudes using
\begin{equation}
S=S(0\,{\rm mag})10^{-0.4(K_{20}+ZPO)}
\end{equation}
where the zero point offset is $ZPO = 0.017\pm0.005$ and $S(0 {\rm
mag})=1.122\times10^{-14}\pm1.891\times10^{-16} {\rm W cm}^{-2}$ for
the $K_{\rm s}$ band (Cohen, Wheaton \& Megeath 2003).

\subsection{The Redshift Distribution and the Selection Function}\label{sec:2mass:nz}

\begin{figure}
\psfig{figure=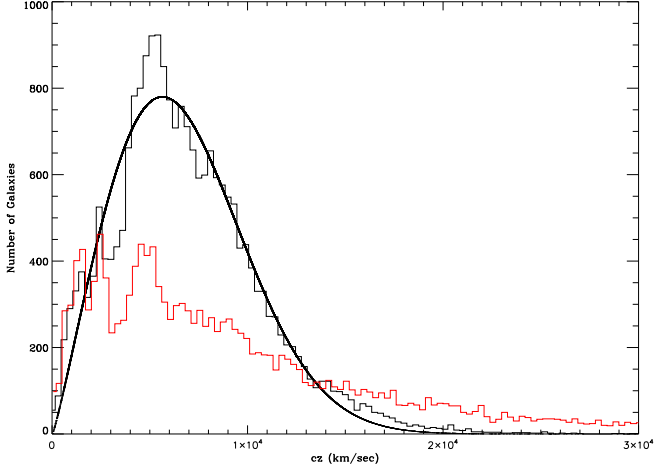,angle=90,width=0.5\textwidth,clip=}
\caption[Redshift histogram for 2MRS galaxies] {Redshift histogram for
2MRS galaxies and a least squares fit (Equation~\ref{eqn:dnz2mrs}) to
the data (black). For comparison, also plotted is a redshift histogram
for PSCz galaxies (Saunders \etal 2000) (red).}
\label{fig:2MRSnz}
\end{figure}

The redshift distribution of the 2MRS is shown in
Figure~\ref{fig:2MRSnz}.  The $IRAS$ PSCz survey redshift distribution
(Saunders \etal 2000) is also plotted for comparison. The 2MRS samples
the galaxy distribution better than the PSCz survey out to
$cz=15000$\kmps. The selection function of the survey (i.e. the
probability of detecting a galaxy as a function of distance) is modeled using a
parametrised fit to the redshift distribution:
\begin{equation}
dN(z)=Az^\gamma\exp\left[-\left(\frac{z}{z_c}\right)^\alpha\right]dz\;,
\label{eqn:dnz2mrs}
\end{equation}
with best-fit parameters of $A=116000\pm4000$, $\alpha=2.108\pm0.003$,
$\gamma=1.125\pm0.025$ and $z_c=0.025\pm0.001$.  This best-fit is also
shown in Figure~\ref{fig:2MRSnz} (solid line). The overall selection
function $\phi(r)$ is the redshift distribution divided by the volume
element 
\begin{equation}
\phi(r)=\frac{1}{\Omega_s r^2} \left(\frac{dN}{dz}\right)_r
\left(\frac{dz}{dr}\right)_r\;
\end{equation}
where $\Omega_s(\approx 4\pi$ steradians) is the solid angle of the
survey and $r$ is the comoving distance.
\subsection{Taking out the Local Group Galaxies}
Galaxies that are members of the Local Group need to be removed
from the 2MRS catalogue to maintain the internal consistency 
of the analysis.
We used the Local Group member list of thirty five galaxies (including Milky Way) 
given in Courteau and Van den Bergh (1999) 
to identify and remove eight LG members (IC 10, NGC 147, NGC 185, NGC 205, NGC 6822, M31, M32, M33) from our analysis.
In Section 4, we will calculate the acceleration on to the Milky Way due to these LG members. 
\subsection{Assigning distances to nearby Galaxies}
In order to reduce the distance conversion errors, we cross-identified 
35 galaxies which have HST Key Project distances (see Freedman \etal 2001 and the references therein) 
and 110 galaxies which have distance measurements compiled from several sources (see Karachentsev \etal 2004 and the references therein). 
We assign these galaxies measured distances instead of converting them from redshifts.
In addition, we identify nine blue-shifted galaxies in the 2MRS which 
are members of the Virgo cluster. 
We assign these galaxies the distance to the centre of Virgo 
($\approx15.4$ Mpc, Freedman \etal 2001). 
Finally, there are four remaining blue-shifted galaxies without known distance measurements which are assigned to 1.18  $h^{-1}$ Mpc
\footnote{This is the zero-velocity
surface which separates the Local Group from the field that is
expanding with the Hubble flow (Courteau \& Van Den Bergh, 1999).}. 
Thus by assigning distances to galaxies, we do not need to 
exclude any non-LG galaxy from the analysis.

\section{The Methods and Weighting Schemes}

In order to compare the CMB and the LG dipoles, it is necessary to
postulate a relation between the galaxy distribution and the
underlying mass distribution. In this paper, we will use both a {\it
number weighted} and a {\it f\mbox{}l\mbox{}ux weighted} prescription.  Although
quite similar in formulation, these schemes are based on very
different models for galaxy formation.  The number weighted
prescription assumes that the mass distribution in the Universe is a
continuous density field and that the galaxies sample this field in a
Poisson way. On the other hand, the f\mbox{}l\mbox{}ux weighted model is based on
the assumption that the mass in the Universe is entirely locked to the
mass of the halos of the luminous galaxies.
\subsection{Number Weighted Dipole}

It is commonly assumed that the galaxy and the mass distributions in
the Universe are directly proportional to each other and are related
by a proportionality constant\footnote{More complicated relations have
been suggested for biasing models, examples include non-linear and
`stochastic' relations. Also, the halo model of clustering involves a
different biasing postulation.}, the linear bias parameter $b$:
$\delta n/n\equiv\delta_g = b \delta_m$.  In this case, Equation~\ref{eqn:v(r)} for
the LG can be rewritten as
\begin{equation}
{\bf v}_{LG}=\frac{H_0 \beta}{4 \pi}\int_{{\bf r}}^{\infty} d^3{\bf
r}^{\prime} \delta_g({{\bf r}^{\prime}}) \frac{{\bf r}^{\prime}-{\bf
r}}{|{\bf r}^{\prime}-{\bf r}|^3}
\label{eqn:v(r)_LG}
\end{equation}
where $\beta \equiv \omegam^{0.6}/b$.

For the number weighted model, incomplete sampling due to the
magnitude limit is described by the selection function, $\phi(r)$
given in Section~\ref{sec:2mass:nz}. Each galaxy $i$ is assigned a
weight:
\begin{equation}
w_i=\frac{1}{\phi(r_i)C_{i}}
\end{equation}
where $\phi(r_i)$ and $C_{i}$ are the values of the radial selection function
and the completeness ($0\leq C_{i} \leq1$) for each galaxy,
respectively.  The observed velocity of the Local Group with respect
to the CMB is given by
\begin{equation}
{\bf v}({\bf r})= \frac{H_0 \beta}{4 \pi \bar{n}}\sum\limits_{i}^N\,
\frac{w_i\hat{{\bf r}}_i}{r_{i}^2}
\label{eqn:obsdip}
\end{equation}
where $\bar{n}$ is the mean galaxy density of the survey and
$\hat{{\bf r}}_i$ is the unit vector of the galaxy's position. The sum
in the equation is over all galaxies in the sample that lie in the
distance range $r_{min}<r_i<r_{max}$.  Calculated this way, the
velocity vector does not depend on the Hubble constant ($h$ cancels
out).  

If the galaxies are assumed to have been drawn by a Poisson point
process from an underlying density field, then it is straightforward
to calculate the shot noise errors.  The shot noise is estimated as
the $rms$ of the cumulative variance, $\sigma_{sn}^2$ given by
\begin{equation}
\sigma_{sn}^2= \left(\frac{H_0 \beta}{4 \pi
\bar{n}}\right)^2\sum\limits_{i}^N\, \left(\frac{\hat{{\bf
r}}_i}{r_{i}^2\phi(r_i)C_{i}}\right)^2\;.
\end{equation}
The shot noise error per dipole component is $\sigma_{1D}=\sigma
_{sn}/ \sqrt{3}$.
\subsection{F\mbox{}l\mbox{}ux Weighted Dipole}

For this model, each galaxy is a `beacon' which represents the
underlying mass. This is characterised by the mass-to-light ratio
$\Upsilon={\rm M}/{\rm L}$.  $\Upsilon$ is probably not constant and
varies with galaxy morphology (e.g. Lanzoni \etal 2004) but
mass-to-light ratios of galaxies vary less in the near-infrared than
in the optical (e.g. Cowie \etal 1994; Bell \& de Jong 2001). In the
context of dipole estimation, this model of galaxy formation implies
that the Newtonian gravitational acceleration vector for a volume
limited sample is \beqn {\bf g}({\bf r})&=& {\rm G} \sum\limits_{i}\,
M_{i} \frac{\hat{{\bf r}}_i}{r_{i}^2} \backsimeq {\rm G}
\left\langle\frac{M}{L}\right\rangle\sum\limits_{i}\, L_{i}
\frac{\hat{{\bf r}}_i}{r_{i}^2}\nonumber \\ &=&4\pi{\rm G}
\left\langle\frac{M}{L}\right\rangle\sum\limits_{i}\, {\rm S}_{i}
\hat{{\bf r}}_i,
\label{eqn:glum}
\eeqn where the sum is over all galaxies in the Universe,
$\left\langle M/L \right\rangle$ is the average mass-to-light ratio
and ${\rm S}_i=L_i/4 \pi r^2$ is the f\mbox{}l\mbox{}ux of galaxy $i$. The peculiar
velocity vector is derived by substituting Equation~\ref{eqn:glum} into the
second line of Equation~\ref{eqn:v(r)}.  For a f\mbox{}l\mbox{}ux limited catalogue
the observed LG velocity is
\begin{equation}
{\bf v}({\bf r})= \frac{8\pi{\rm G}f(\Omega_{\rm m})}{3 H_0\Omega_{\rm
m}b_{\rm L}} \left\langle\frac{M}{L}\right\rangle\sum\limits_{i}\,
w_{L_i}{\rm S}_{i} \hat{{\bf r}}_i
\label{eqn:vl}
\end{equation}
where $b_{\rm L}$ is the luminosity bias factor introduced to account for the
dark matter haloes not fully represented by 2MRS galaxies and $w_{L_i}$
is the weight assigned to galaxy $i$ derived in the next section.  The
mass-to-light ratio, assumed as constant, is given by
\begin{equation}
\left\langle\frac{M}{L}\right\rangle = \frac{\rho_{\rm m}}{\rho_{\rm
L}} =\frac{3 H_0^2 \Omega_{\rm m}}{8\pi G \rho_{\rm L}}
\end{equation} 
where $\rho_{\rm L}$ is the luminosity density so
Equation~\ref{eqn:vl} is rewritten as:
\begin{equation}
{\bf v}({\bf r})= \frac{H_0f(\Omega_{\rm m})}{\rho_{\rm L}b_{\rm
L}}\sum\limits_{i}^{N}\, w_{L_i}{\rm S}_{i} \hat{{\bf r}}_i.
\label{eqn:vlum}
\end{equation}

The flux weighting method (originally proposed by Gott) has been applied extensively to 
two-dimensional galaxy catalogues 
(e.g. Yahil, Walker \& Rowan-Robinson 1986, Villumsen \& Strauss 1987, 
Lahav, Rowan-Robinson \& Lynden-Bell 1988) and most recently to 2MASS XSC (Maller \etal 2003). 
Since these surveys lack radial information, the dipoles were calculated by 
either assuming $w_{L_i}=1$ (e.g. Maller \etal 2003) or by using a luminosity function 
based on a redshift survey in a section of the two-dimensional catalogue 
(e.g. Lahav, Rowan-Robinson \& Lynden-Bell 1988). In either case, it was not possible to 
determine the convergence of the dipole as a function of redshift. 
The three-dimensional $IRAS$ dipoles (Strauss \etal 1990,  Webster, Lahav \& Fisher 1997, Schmoldt \etal 1999 and Rowan-Robinson \etal 2000) 
were derived using the number weighted
scheme because the $IRAS$ catalogues are biased towards star
forming galaxies with widely varying mass-to-light ratios 
resulting in a very broad luminosity function making it difficult to 
estimate the distance. The 2MRS is mainly 
sensitive to total stellar mass rather than instantaneous star formation 
rates (e.g. Cole et al. 2001) 
and consequently the 2MRS mass-to-light ratios do not have
as much scatter. Thus, for the first time, 
the 2MRS enables the determination the convergence of the flux dipole
as a function of distance.
There are many advantages to using the f\mbox{}l\mbox{}ux weighted
model for the dipole calculation and these will be discussed in the
coming sections.

\begin{figure}
\psfig{figure=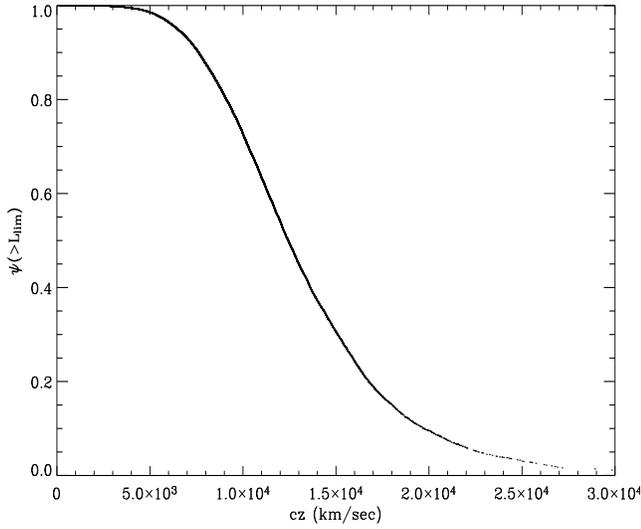,angle=0,width=0.5\textwidth,
height=75mm, clip=}
\caption[F\mbox{}l\mbox{}ux Weighted Selection Function of 2MRS] {The f\mbox{}l\mbox{}ux weighted
selection function as a function of redshift.}
\label{fig:lum}
\end{figure}

For the f\mbox{}l\mbox{}ux weighted case, the weighting function is derived as
follows: Let $\rho_{\rm L}({\rm L}\geq 0)$ be the luminosity density
in the volume element $\delta V$ of a volume limited catalogue. In
this case the dipole velocity is simply
\begin{equation}
{\bf v}({\bf r})= \frac{H_0f(\Omega_{\rm m})}{\rho_{\rm L}b_{\rm
L}}\sum\limits_{i}^{N}\, \frac{ \delta V_i \rho_{\rm L}({\rm L_i}\geq
0) \hat{{\bf r}}_i}{r_i^2}.
\end{equation}
In practice, however, we have a f\mbox{}l\mbox{}ux limited catalogue with $S \geq
S_{\rm lim}$ so only galaxies with luminosity $L \geq L_{\rm lim}=4
\pi r^2 S_{\rm lim}$ are included in the survey. Thus the total
luminosity in the infinitesimal volume $\delta V$ is
\begin{equation} 
\delta V \rho_{\rm L}({\rm L}\geq 0)=L_{\rm obs}+\delta V \rho_{\rm
L}({\rm L}< {\rm L}_{\rm lim})
\label{eqn:dv}
\end{equation}
where $L_{\rm obs}=\delta V \rho_{\rm L}({\rm L} \geq {\rm L}_{\rm
lim})$ is the observed luminosity and $\delta V \rho_{\rm L}({\rm L} <
{\rm L}_{\rm lim})$ is the luminosity that was not observed due to the
f\mbox{}l\mbox{}ux limit of the survey. Substituting
\begin{equation}
\delta V= \frac{L_{\rm obs}}{\rho_{\rm L}({\rm L} \geq {\rm L}_{\rm
lim})}
\end{equation}
into Equation~\ref{eqn:dv} yields \beqn \delta V \rho_{\rm L}({\rm
L}\geq 0)& = & L_{\rm obs}\Bigg[1+\frac{\rho_{\rm L}({\rm L}< {\rm
L}_{\rm lim})}{\rho_{\rm L}({\rm L}\geq {\rm L}_{\rm lim})}\Bigg]
\nonumber \\ & = &L_{\rm obs}\Bigg[1+\frac{\rho_{\rm L}({\rm L}\geq
0)-\rho_{\rm L}({\rm L}\geq {\rm L}_{\rm lim})}{\rho_{\rm L}({\rm
L}\geq {\rm L}_{\rm lim})}\Bigg] \nonumber \\ & = &L_{\rm
obs}\frac{\rho_{\rm L}({\rm L}\geq 0)} {\rho_{\rm L}({\rm L}\geq {\rm
L}_{\rm lim})}\equiv \frac{L_{\rm obs}} {\psi({\rm L}\geq {\rm L}_{\rm
lim})}
\label{eqn:phiL}
\eeqn where $\psi({\rm L}\geq {\rm L}_{\rm lim})$ is the f\mbox{}l\mbox{}ux weighted
selection function. In Figure~\ref{fig:lum}, the interpolated fit
$\psi({\rm L}\geq {\rm L}_{\rm lim})$ for the 2MRS galaxies is shown
as a function of redshift.

Thus, the overall weight factor, $w_L$, is
\begin{equation}
w_{L_i}=\frac{1}{\psi(r_i)C_{i}}.
\end{equation}
The luminosity density of the 2MRS is
\begin{equation}
\rho_L=\frac{1}{V}\sum_i\; \frac{L_i}{\psi(L\geq
L_{i,lim})}=(7.67\pm1.02)\times10^8 \rm\thinspace L_{\odot} h {\rm
Mpc}^{-3}
\end{equation}
where $V$ is the survey volume. The value of $\rho_L$ is in good
agreement with the value derived by Kochanek \etal (2001),
$\rho_L=(7.14\pm0.75)\times10^8$ \Lsun $h {\rm Mpc}^{-3}$.
We note that the number weighted selection function, $\phi(r)$, drops 
with distance faster than the luminosity weighted selection function, 
$\psi(r)$.
At large distances, we observe only the most luminous galaxies, so  
the amount of `missing' luminosity from a volume of space is not as 
big as the number of `missing' galaxies. Therefore, as shown below, 
the f\mbox{}l\mbox{}ux weighted dipole is more robust at large distances than the number weighted dipole.  

For the f\mbox{}l\mbox{}ux weighted scheme, the shot noise is estimated as
\begin{equation}
\sigma_{sn}^2= \left(\frac{H_0f(\Omega_{\rm m})}{\rho_{\rm
L}}\right)^2 \sum\limits_{i}\, \left(\frac{S_i\hat{{\bf
r}}_i}{\psi(r_i)C_{i}}\right)^2\;.
\end{equation}
As the exact shot noise effects for the different models of galaxy
formation are difficult to model, we will use the Poisson estimate
above as an indicator of uncertainties (see also Kaiser \& Lahav
1989). However, we note that the quoted uncertainties overestimate the
noise at small distances where the survey is volume limited and
underestimate it at large distances where only the brightest galaxies
are sampled.  The uncertainties and the variation in the mass-to-light
ratios relation also affect the calculation but they are not accounted
for in this analysis.
\subsection{The Redshift-Space Effects}

It is well known that the peculiar velocities distort the pattern of
density enhancements in redshift-space.  The peculiar acceleration of
the Local Group calculated using redshifts instead of real distances
will differ from the actual LG acceleration (Kaiser 1987 and Kaiser
\& Lahav 1989). This effect, referred to as {\it the rocket effect},
is easily visualised by supposing that only the Local Group has a velocity in
an homogeneous universe without any peculiar velocities.  If the LG
frame redshifts are used as distance indicators then there will be a
spurious contribution from the galaxies that are in the direction of
the LG motion. The prediction for net spurious acceleration is given
by (Kaiser \& Lahav 1988):
\begin{equation}
{\bf v}_{spur}=\frac{1}{3}{\bf v}({\bf 0})\Bigg[
\Bigg(2+\frac{d\ln\phi}{d\ln r}\Bigg)_{z_{vol}}
\ln\frac{z_{vol}}{z_{min}}+\ln\frac{\phi(z_{max})z_{max}^2}
{\phi(z_{vol})z_{vol}^2}\Bigg]
\end{equation}
where $z_{min}$ is the minimum, $z_{max}$ is the maximum redshift of
the survey and $z_{vol}$ is the redshift for which the survey is
volume limited ($cz_{vol}\approx$4500 \kmps for the 2MRS).

The rocket effect is very important for the number weighted LG dipole
calculation because of the dependence on $r_i$ in
Equation~\ref{eqn:obsdip}.  For the 2MRS, the predicted error from the rocket effect is a contribution to the total dipole by
roughly ${\bf v}_{spur}\approx-0.6{\bf v}({\bf 0})$.  There are two
ways to overcome this error. One is to work in real-space instead of
redshift-space. This will be discussed in a forthcoming paper where
the Local Group dipole will be calculated using the Wiener
reconstructed real-space density field. The other one is to use the
f\mbox{}l\mbox{}ux weighted model. For the f\mbox{}l\mbox{}ux weighted LG dipole, the rocket
effect is almost negligible as it plays a role only in the
determination of the radii of the concentric spheres within which the
dipole is calculated.

\subsection{The Reference Frames}

Brunozzi \etal (1995) and Kocevski, Mullis \& Ebeling (2004) 
claim that the
dipole in the LG and the CMB frames are over- and under-estimates of
the real dipole, respectively. The redshift of an object is defined by
\begin{equation}
cz=H_0r+({\bf v}({\bf r})-{\bf v}({\bf 0}))\cdot\hat{{\bf r}}
\end{equation} 
where ${\bf v}({\bf r})$ is the peculiar velocity of the object and
${\bf v}({\bf 0})$ is the observer's peculiar velocity. In the LG
frame $|{\bf v}({\bf 0})|=627\pm22$\kmps and in the CMB frame $|{\bf
v}({\bf 0})|=0$\kmps by definition.  Therefore, the redshift of a
galaxy that has the same direction of motion as the LG would be larger
in the CMB frame than that in the LG frame. In this case, since the
acceleration vector is proportional to $1/r^2$, the amplitude of the 
dipole in the LG frame is expected to be larger than the amplitude
of the dipole in the CMB frame. As the dipole is thought to be
dominated by the nearby objects that participate together with the LG
in a bulk motion (i.e. ${\bf v}({\bf r})\approx{\bf v}({\bf 0})$ so
that $cz_{LG}\approx H_0r$), it is often assumed that the real LG
dipole is closer to the dipole in the LG frame than that in the CMB
frame. On the other hand, Branchini \& Plionis (1996) find that the real-space
reconstruction of the LG dipole gives a result halfway between the LG
frame and the CMB frame values.

We perform the analysis using both the LG and the CMB frame redshifts.
All galaxies are referenced to the rest frame of the LG using the
transformation in consistency with who use the same conversion 
Courteau \& Van Den Bergh (1999):
\begin{equation}
cz_{LG}=cz_{hel}-79\cos(l)\cos(b)+296\sin(l)\cos(b)-36\sin(b)
\end{equation}
where $z_{hel}$ is the heliocentric redshift and $l$ and $b$ are the
longitude and the latitude of the galaxy in the Galactic coordinates,
respectively. We convert from the LG frame to the CMB frame using  
\beqn
cz_{CMB}&=&cz_{LG}+v_{LG}[\sin(b)\sin(b_{LG}) \nonumber \\
&+&\cos(b)\cos(b_{LG})\cos(|l_{LG}-l|)], \eeqn where $v_{LG}$ is the
amplitude of the LG velocity with respect to the CMB and ($l_{LG}$,
$b_{LG}$) is the direction of its motion.  We use the CMB dipole value
of Bennett \etal (2003). Using the first year of data from WMAP, they
find that the Sun is moving at a speed of 369.5$\pm$3.0 km ${\rm
s}^{-1}$, towards ($l=263.85^\circ\pm0.10^\circ$,
$b={48.25}^\circ\pm0.40^\circ$). Using the revised values of
the motion of the Sun relative to the LG a velocity of 306$\pm$18 \kmps towards ($l=99^\circ\pm5^\circ$,
$b={-4}^\circ\pm4^\circ$) derived by Courteau \& Van Den Bergh
(1999), we find a LG velocity relative to the CMB of $v_{LG}=627\pm22$
\kmps, towards ($l_{LG}=273^\circ\pm3^\circ$,
$b_{LG}=29^\circ\pm3^\circ$).

The choice of reference frames highlights another advantage of the
f\mbox{}l\mbox{}ux weighted dipole calculation. As the redshifts only enter the
calculation in the determination of the radius of the concentric
spheres, the results are robust to changes in reference frames.
\section{Dipole Results}\label{sec:dip:results}

\begin{figure*}
\psfig{figure=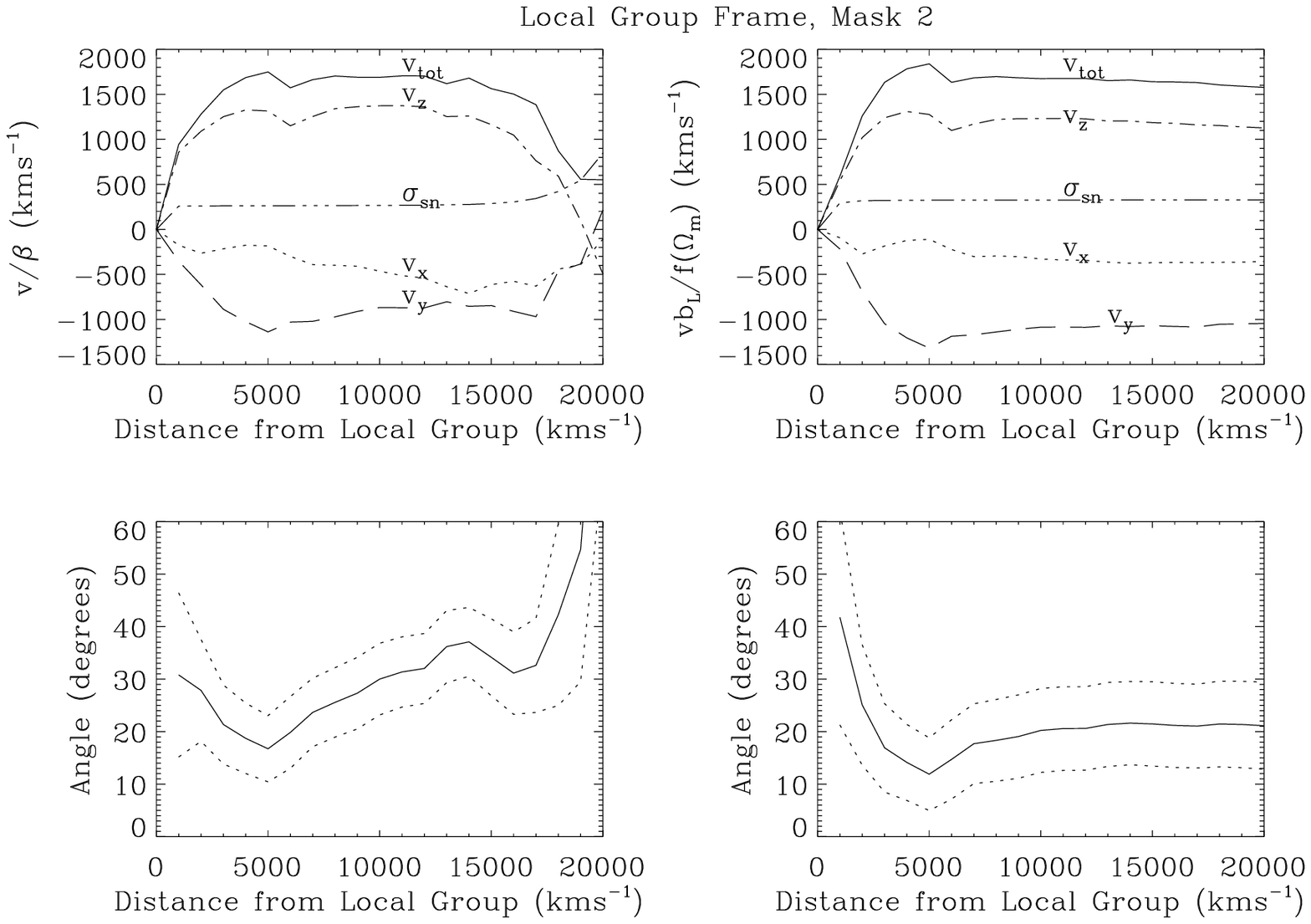,angle=0.,height=100mm, 
width=\textwidth,clip=}
\caption[The convergence of the LG acceleration in LG frame (Mask 2)]
{{\bf Top}: Three components and the magnitudes of the acceleration of the
Local Group due to galaxies within a series of successively larger
concentric spheres centred on the local group in the {\bf Local Group
frame}. The galaxies in the
masked regions are interpolated from the adjacent
regions (Method 2). 
{\bf Left} panel is {\bf the
number weighted} velocity and {\bf right} panel is {\bf the f\mbox{}l\mbox{}ux
weighted} velocity. The growth of the estimated shot noise is also
shown. {\bf Bottom}:
Convergence of the direction of the LG dipole where the misalignment
angle is between the LG and the CMB dipoles. The dotted lines denote
1$\sigma$ errors from shot noise. Left plot is the direction of the number
weighted LG dipole and right plot is the direction of the f\mbox{}l\mbox{}ux
weighted LG dipole. We note for the number weighted dipole the dramatic increase in shot noise beyond
15000 \kmps, where the dipole's behaviour cannot be interpreted reliably.}
\label{fig:velvsdist}
\psfig{figure=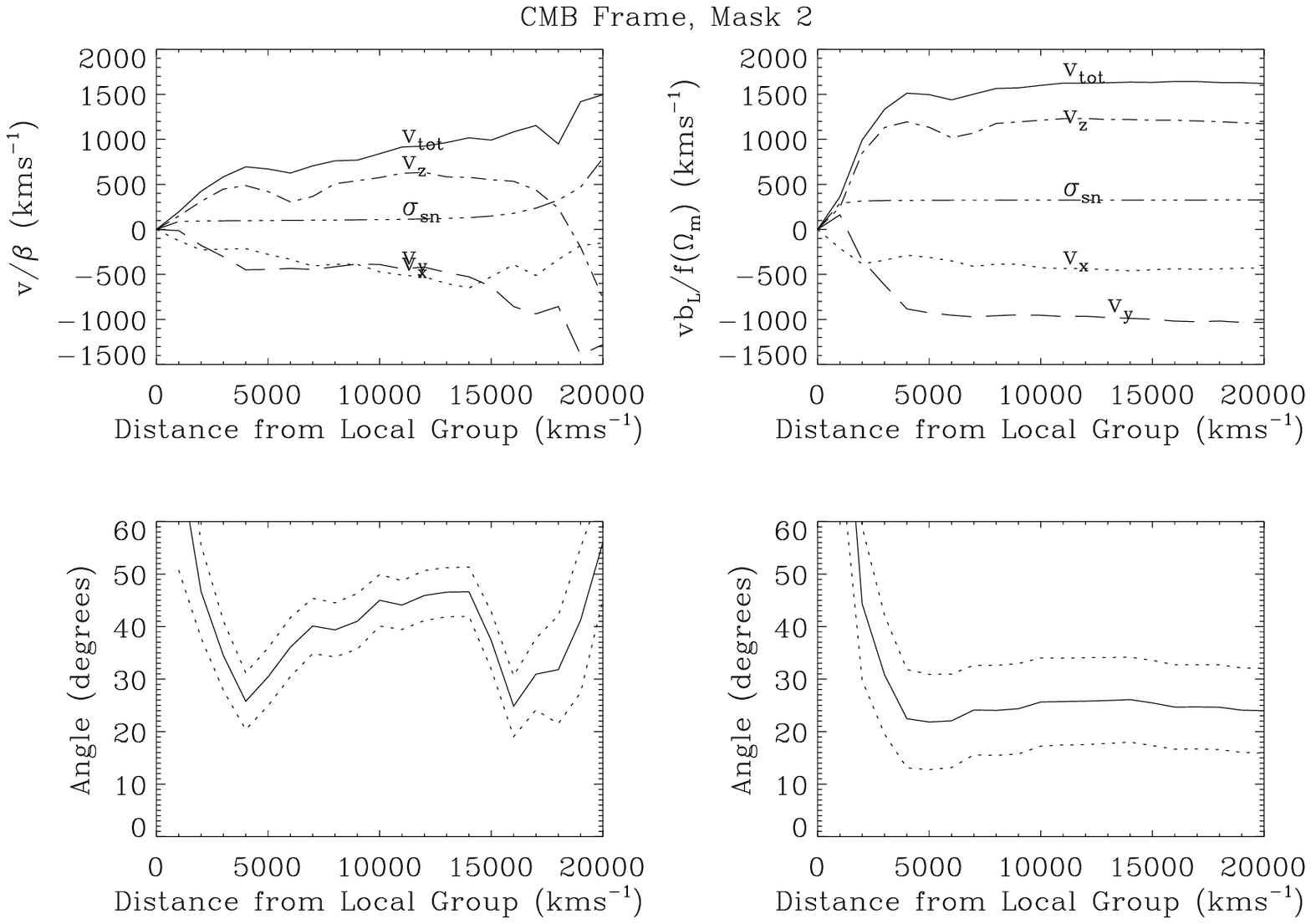,angle=0.,height=100mm, 
width=\textwidth,clip=}
\caption[The convergence of the LG acceleration in CMB frame (Mask 2)]
{Same as in Figure~\ref{fig:velvsdist} but in {\bf CMB frame}.}
\label{fig:velvsdistcmb}
\end{figure*}

\begin{figure*}
\psfig{figure=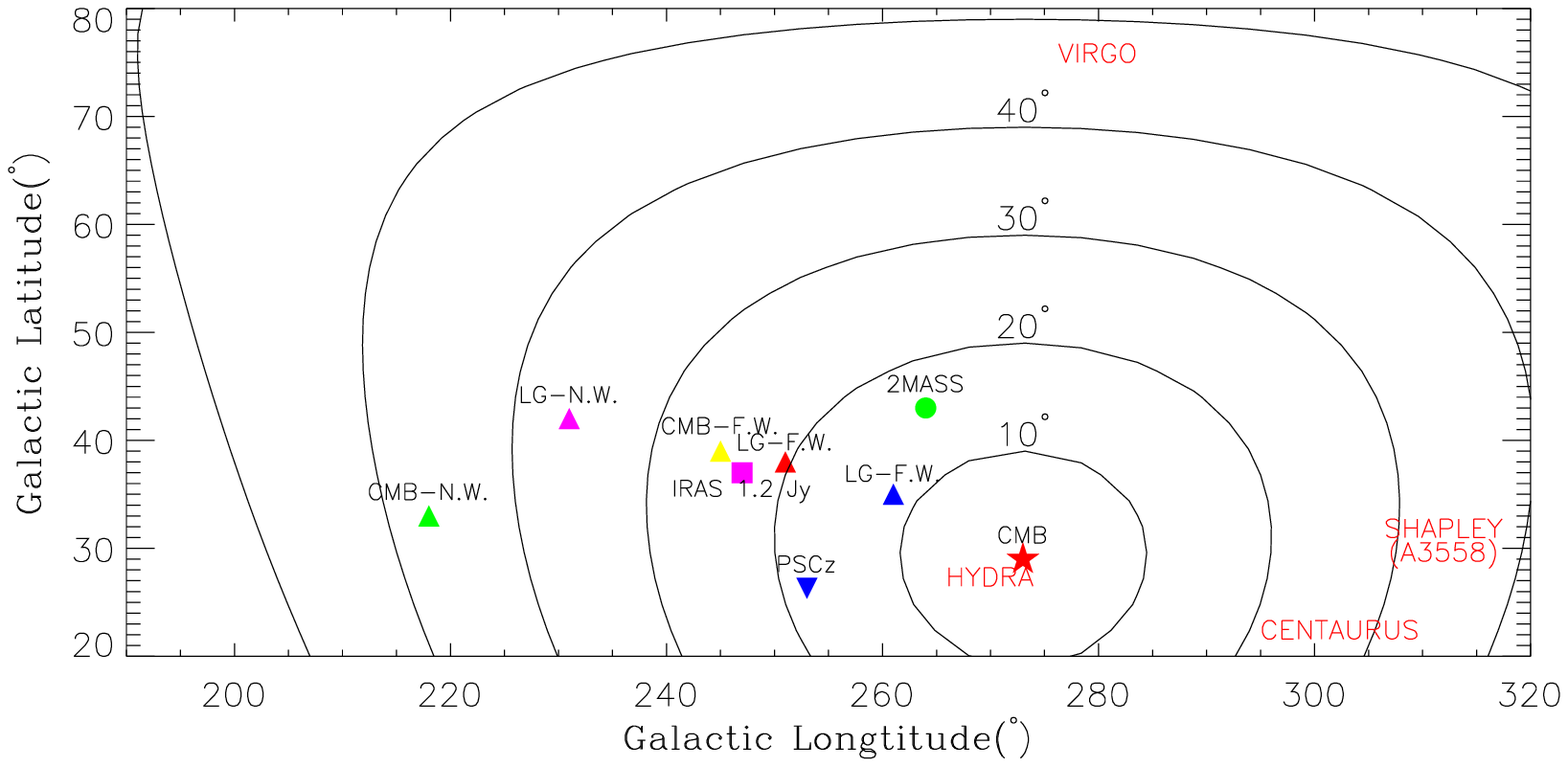,angle=0.,width=\textwidth,clip=}
\caption[]{The triangles show
the dipole directions at 13000 \kmps derived in this paper.
The direction of the number weighted 2MRS dipole in the LG frame is
shown in magenta (LG-N.W., $l=231^\circ$, $b=42^\circ$). The red
triangle shows the direction of the f\mbox{}l\mbox{}ux weighted 2MRS in the LG frame
(LG-F.W., $l=251^\circ$, $b=38^\circ$). 
The blue
triangle shows the direction of the f\mbox{}l\mbox{}ux weighted 2MRS in the LG frame excluding Maffei 1, Maffei 2, M81, IC342 and Dwingeloo 1
(LG-F.W., $l=261^\circ$, $b=35^\circ$). The red star shows the CMB
dipole direction ($l=273^\circ$, $b=29^\circ$). The green triangle is
the number weighted LG dipole in the CMB frame (CMB-N.W., $l=218^\circ$,
$b=33^\circ$); the yellow triangle is the f\mbox{}l\mbox{}ux weighted LG dipole in
CMB frame (CMB-F.W., $l=245^\circ$, $b=39^\circ$).  
The green circle is the 2MASS dipole
($l=264^\circ$, $b=43^\circ$, Maller \etal 2003).  The magenta square
is the $IRAS$ 1.2-Jy dipole ($l=247^\circ$, $b=37^\circ$, Webster
\etal 1997). The blue upside-down triangle is the $IRAS$ PSCz dipole
($l=253^\circ$, $b=26^\circ$, Rowan-Robinson \etal 2000). Contours are
drawn at constant misalignment angles. Also shown are the Virgo
Cluster ($l=280^\circ$, $b=75^\circ$), the Hydra Cluster
($l=270^\circ$, $b=27^\circ$), the Centaurus Cluster ($l=302^\circ$,
$b=22^\circ$) and A3558 ($l=312^\circ$, $b=31^\circ$).}
\label{fig:directions}
\end{figure*}

The results are presented in
Figures~\ref{fig:velvsdist}-\ref{fig:directions}.  The top plots of
Figures~\ref{fig:velvsdist}-\ref{fig:velvsdistcmb} show the
amplitudes and three spatial components of the acceleration on the
Local group (top) and the convergence of the angle between the LG
dipole and the CMB dipole (bottom) as a function of distance in two 
reference frames.  For these plots, the galaxies in the
masked regions are interpolated from the adjacent
regions (Method 2). 
The right panel in each figure shows the results for the f\mbox{}l\mbox{}ux weighted dipole and the left panels show the results for the number weighted
dipole. Figure~\ref{fig:velvsdist} is for the 
the Local Group Frame and Figure~\ref{fig:velvsdistcmb} is  
in the CMB frame. As discussed in the next section, the results 
for the filling Method 1 where the galaxies are sampled randomly 
do not look very different 
than the results in Figures~\ref{fig:velvsdist} \& \ref{fig:velvsdistcmb} and 
thus are not shown. 
Figure~\ref{fig:directions} compares the direction
of the LG dipole estimate to that of the CMB and other LG dipole
measurements.  

We give the results for the f\mbox{}l\mbox{}ux weighted dipole calculated using
the second method of mask filling in Table~\ref{tab:tab2}.  Column 1
is the radii of the concentric spheres within which the values are
calculated; columns 2 and 3 are the amplitude of the velocity vector
and the shot noise divided by $\omegam^{0.6}/b_{\rm L}$, respectively;
Columns 4, 5 and 6 show the direction of the velocity vector and its
angle to the CMB dipole. The first line gives the results in the LG
frame and the second line gives the results in CMB frame. Table~\ref{tab:tab3} is structured 
in the same way as Table~\ref{tab:tab2} however the analysis excludes five galaxies.

\subsection{The Tug of War}
In Figures~\ref{fig:velvsdist} \& \ref{fig:velvsdistcmb}, 
the LG velocity is dominated by structure within a
distance of 60 \kmps (except for the CMB frame number weighted
dipole where the contribution from the distant structure is over-estimated.). 
The `tug of war' between the Great
Attractor and the Perseus-Pisces is clearly evident. The dip in the
velocity vector is an indication that the local flow towards the Great
Attractor\footnote{ By `Great Attractor', it is meant the entire
steradian on the sky centred at ($l\sim310^\circ$,$b\sim20^\circ$)
covering a distance of 20 $h^{-1}$ Mpc to 60 $h^{-1}$ Mpc.} is
counteracted by the Perseus-Pisces complex in the opposite
direction. If we take out 420 galaxies in the Perseus-Pisces ridge
(defined by $-40 \le b \le -10$, $110 \le b \le 130$, 4600 ${\rm km s}^{-1}
\le cz \le$ 6000 ${\rm km s}^{-1}$) and recalculate the convergence,
the dip almost disappears and the convergence is dominated by the
Great Attractor. This leads us to conclude that the Perseus-Pisces
plays a significant role in the gravitational acceleration of the LG.

\subsection{Filling the Zone of Avoidance}
The choice of method used to fill the masked regions does not have
much effect on the results. The convergence of the misalignment angle
for the second method is slightly more stable than the first method
and the overall direction of the LG dipole is closer to the CMB
dipole. Since the Galactic $z$ component is least affected by the zone
of avoidance, the discrepancy between the amplitudes in each plot
comes mainly from the Galactic $x$ and $y$ components. The direction
of the dipole is 2$^\circ$-3$^\circ$ closer to the CMB vector for the
second method at distances where the Great Attractor lies. 
Of course, one
cannot rule out the possibility that there may be important
contribution to the dipole from other structures behind the zone of
avoidance. The Great Attractor is most likely centred on the Norma Cluster ($l\approx$325$^\circ$, $b\approx$-7$^\circ$, Kraan-Korteweg \etal 1996) which lies very close to the obscured plane. Although, the 2MRS samples the Norma cluster much better than the optical surveys, 
the latitude range $|b|\lesssim5$ which is still obscured in the 2MRS 
may have structure that play an important role in the dipole determinations. 
In fact, Scharf \etal (1992) and Lahav \etal (1993) point
out that there is significant contribution to the local flow by the
Puppis complex at low galactic latitudes that are not sampled by the 2MRS.   

Kraan-Korteweg \& Lahav (2000) point out that  
since the dipole is dominated by local structures, the detection of nearby galaxies can be more 
important to the dipole analyses 
than the detection of massive clusters at larger distances. 
We test this by excluding the five most luminous nearby galaxies (Maffei 1, Maffei 2, IC342, Dwingeloo 1 and M81) from our analysis. 
Remarkably, the direction of the resultant dipole moves much closer 
to that of the CMB (see Table~\ref{tab:tab3}). 
All of these galaxies expect M81 lie very close to the Zone of Avoidance and 
they are excluded from most dipole analyses either because they are not in the catalogue (e.g. Rowan-Robinson \etal 2000) or 
they are masked out (e.g. Maller \etal 2003). In fact, when we change our mask to match that of Maller \etal (2003) and keep M81 in the analysis our resulting dipole is only 3$^\circ$ degrees away from the dipole calculated by Maller \etal (2003).
This illustrates the importance of the nearby 
structure behind the Zone of Avoidance.

The comparison of Tables~\ref{tab:tab2} and~\ref{tab:tab3} also highlights the vital role non-linear dynamics induced by 
nearby objects play in dipole calculations.    
Maller \etal (2003) investigate the non-linear contribution to the LG dipole by removing the bright 
galaxies with $K_s<8$. 
They report that the LG dipole moves to within a few degrees of the CMB dipole. They repeat 
their calculations for the PSCz survey and observe the same pattern. We do not 
observe this behaviour. When we remove the objects brighter that $K_s=8$ 
(428 galaxies), the misalignment angle of the resulting dipole decreases by 
$3^\circ$ for the flux weighted dipole and remains the same for 
the number weighted case. The dipole 
amplitudes decreases substantially 
in both cases, notably in the case of the flux dipole, suggesting 
that the brightest 2MRS galaxies play a significant role in inducing the LG
velocity.

\subsection{The Choice of Reference Frames}
The number weighted LG dipole looks very different in different
reference frames (Figures~\ref{fig:velvsdist}
and~\ref{fig:velvsdistcmb}) whereas the f\mbox{}l\mbox{}ux weighted dipole is almost
unaffected by the change. 
The number weighted dipole is similar to the 
f\mbox{}l\mbox{}ux weighted dipole in the LG frame. 
Thus, we conclude that it is more accurate to use the LG
frame redshifts than that of the CMB frame.

\subsection{The Choice of Weighting Schemes}
The amplitudes of the number weighted LG dipole and the f\mbox{}l\mbox{}ux weighted
LG dipole are very similar in the LG frame. However, the convergence
of the misalignment angles of the flux and the number dipoles 
is quite different in both frames, especially at large distances.  The
angle of the f\mbox{}l\mbox{}ux weighted dipole is closer to the CMB
dipole than its number weighted counterpart at all distances. 
With either weighting
scheme, the dipoles are closest to the CMB dipole at a
distance of about 5000 \kmps and move away from the CMB
dipole direction further away. However, the change in the direction of
the f\mbox{}l\mbox{}ux weighted dipole is much smaller compared to the number
weighted dipole and there is convergence within the error bars by 6000 \kmps. The misalignment angles in the LG frame 
at 130 $h^{-1}$ Mpc are 21$^\circ$ and $37^\circ$ for the flux and the 
number dipoles, respectively. 
The discrepancy is probably 
mainly due to the fact the number dipole is plagued with errors due to the lack of peculiar velocity information.
In fact, when we use just the redshift information instead of the distance measurements (see Section 2.4) the number dipole moves $\approx7^\circ$ towards the CMB dipole. 
The flux dipole assumes that the mass traces light whereas the number dipole assumes that all galaxies have the same mass. The 
former assumption is of course more valid, however, since the amplitudes of the dipoles are so similar in the LG frame, 
we conclude that the 
equal mass assumption for the number weighted dipole 
do not introduce large errors and that the discrepancy results from the errors in distance. 

In all figures, $v_x$ and $v_y$ change with distance more rapidly in the number weighted scheme than the f\mbox{}l\mbox{}ux weighted. 
At further distances, the f\mbox{}l\mbox{}ux weighted
$v_x$ flattens whereas its number weighted counterpart continues to
grow. 
It is expected that the $(x,y)$ directions are particularly 
sensitive to the shape of the zone of avoidance, although it is not obvious why the f\mbox{}l\mbox{}ux weighted components remain so robust. 
Assuming the dipole has converged we can obtain values for 
$\omegam^{0.6}/b$ (number weighted) and $\omegam^{0.6}/b_{\rm L}$ (f\mbox{}l\mbox{}ux weighted) by
comparing the amplitude of our dipole estimates to the CMB dipole.
These values are summarised in
Table~\ref{tab:omegavebeta}. The values are quoted in the LG frame at\footnote{This is the distance beyond which
the shot noise becomes too high (over 10$\%$ for the number weighted
analysis).}
13000 \kmps using the second mask and with the luminosity
density value derived earlier, $\rho_L=7.67\pm1.02\times10^8$ \Lsun $h
{\rm Mpc}^{-3}$.  The errors take the shot noise, the uncertainties in
the CMB dipole and $\rho_L$ (for the f\mbox{}l\mbox{}ux limited case) into account. The $\beta$ values obtained for the two different weighting schemes are in excellent agreement suggesting that the dark matter haloes are well sampled by the survey. Our value for $\beta$ is also in good agreement with results from 2MASS (Pike \& Hudson 2005) and IRAS surveys (e.g. Zaroubi \etal 2002, Willick \& Strauss 1998).
In order to calculate the uncertainties introduced by the errors in
the galaxy redshifts, ten realisations of the 2MRS catalogue are
created with each galaxy redshift drawn from a Gaussian distribution
with standard deviation equal to its error\footnote {The mean value of
the redshift measurement errors is 30 \kmps.}. 
It is found that the scatter in the
dipole results due to the errors in redshifts are very small compared
to the shot noise errors and thus are not quoted. 

\begin{table}
\caption[The values derived for $\omegam$, $\beta$] {The values
derived for $\omegam$, $\beta$.}
\begin{center}
\begin{tabular}{|l|c|c|}
\hline
$\omegam^{0.6}/b_{\rm L}$ from the f\mbox{}l\mbox{}ux weighted scheme& $=$ &
$0.40\pm0.09$ \\\hline $\omegam^{0.6}/b$ from the number weighted scheme& $=$
& $0.40\pm0.08$ \\ \hline
\end{tabular}
\end{center}
\label{tab:omegavebeta}
\end{table}

\subsection{The Milky Way Dipole}
We also investigate the acceleration on our galaxy due to other eight members of the LG excluded from the LG dipole analysis. 
As expected, the flux dipole is strongly dominated by Andromeda (M31) 
with an amplitude of ${\rm v}/\beta\approx220$\kmps directly towards M31 ($l\approx121.4^\circ, b\approx-21.7^\circ$), confirming that near-infrared fluxes are good tracers of mass. 
The number weighted dipole which assumes that the galaxies have the same weight gives a similar amplitude of ${\rm v}/\beta\approx190$\kmps but its direction ($l\approx104.6^\circ, b\approx-21.6^\circ$) is skewed towards NGC 6822 ($l\approx25.3^\circ, b\approx-18.4^\circ$) which lies further away from 
the other seven galaxies that are grouped together.

\begin{table*}
\caption[The dipole convergence values for f\mbox{}l\mbox{}ux weighted dipole] {The
convergence values calculated for f\mbox{}l\mbox{}ux weighted dipole using the
second method for the masked region. The columns give (i)Radius of the
concentric spheres within which the values are calculated; (ii) and
(iii) amplitude of the velocity vector and the shot noise divided by
$f(\Omega_{\rm m})$, respectively; (iv), (v) and (vi) direction of the
velocity vector and its angle to the CMB dipole. The results in the
first and second line are given in LG and CMB frames, respectively.}
\begin{center}
\begin{tabular}{@{}ccccc}
\hline \\ Dist & $V_{tot}b_L/f(\Omega_{\rm m})$ & l & b &
$\delta\theta$ \\ km $s^{-1}$ & km $s^{-1}$ & deg & deg & deg \\ \hline \\
 1000  &  590      $\pm$294          &259$^\circ\pm$ 59  $^\circ$& 39$^\circ\pm$20  $^\circ$& 42$^\circ\pm$20 $^\circ$\\
     &  362      $\pm$289          &331$^\circ\pm$171  $^\circ$& 23$^\circ\pm$23  $^\circ$& 99$^\circ\pm$26 $^\circ$\\\hline
 2000  & 1260      $\pm$318          &249$^\circ\pm$ 17  $^\circ$& 41$^\circ\pm$13  $^\circ$& 25$^\circ\pm$11 $^\circ$\\
     &  993      $\pm$316          &222$^\circ\pm$ 22  $^\circ$& 41$^\circ\pm$15  $^\circ$& 44$^\circ\pm$15 $^\circ$\\\hline
 3000  & 1633      $\pm$322          &260$^\circ\pm$ 12  $^\circ$& 40$^\circ\pm$10  $^\circ$& 17$^\circ\pm$ 8 $^\circ$\\
     & 1334      $\pm$320          &241$^\circ\pm$ 17  $^\circ$& 44$^\circ\pm$13  $^\circ$& 31$^\circ\pm$11 $^\circ$\\\hline
 4000  & 1784      $\pm$323          &264$^\circ\pm$ 10  $^\circ$& 39$^\circ\pm$ 9  $^\circ$& 14$^\circ\pm$ 7 $^\circ$\\
     & 1513      $\pm$322          &252$^\circ\pm$ 14  $^\circ$& 41$^\circ\pm$11  $^\circ$& 22$^\circ\pm$ 9 $^\circ$\\\hline
 5000  & 1838      $\pm$324          &265$^\circ\pm$ 10  $^\circ$& 36$^\circ\pm$ 9  $^\circ$& 12$^\circ\pm$ 7 $^\circ$\\
     & 1497      $\pm$323          &252$^\circ\pm$ 13  $^\circ$& 39$^\circ\pm$11  $^\circ$& 22$^\circ\pm$ 9 $^\circ$\\\hline
 6000  & 1633      $\pm$325          &259$^\circ\pm$ 10  $^\circ$& 34$^\circ\pm$ 9  $^\circ$& 15$^\circ\pm$ 8 $^\circ$\\
     & 1438      $\pm$324          &250$^\circ\pm$ 12  $^\circ$& 36$^\circ\pm$11  $^\circ$& 22$^\circ\pm$ 9 $^\circ$\\\hline
 7000  & 1682      $\pm$325          &256$^\circ\pm$ 10  $^\circ$& 36$^\circ\pm$ 9  $^\circ$& 18$^\circ\pm$ 8 $^\circ$\\
     & 1503      $\pm$324          &247$^\circ\pm$ 12  $^\circ$& 36$^\circ\pm$10  $^\circ$& 24$^\circ\pm$ 9 $^\circ$\\\hline
 8000  & 1697      $\pm$326          &255$^\circ\pm$ 11  $^\circ$& 37$^\circ\pm$10  $^\circ$& 18$^\circ\pm$ 8 $^\circ$\\
     & 1566      $\pm$325          &248$^\circ\pm$ 12  $^\circ$& 39$^\circ\pm$10  $^\circ$& 24$^\circ\pm$ 9 $^\circ$\\\hline
 9000  & 1683      $\pm$326          &255$^\circ\pm$ 11  $^\circ$& 38$^\circ\pm$10  $^\circ$& 19$^\circ\pm$ 8 $^\circ$\\
     & 1573      $\pm$325          &248$^\circ\pm$ 12  $^\circ$& 39$^\circ\pm$10  $^\circ$& 24$^\circ\pm$ 9 $^\circ$\\\hline
10000  & 1674      $\pm$326          &253$^\circ\pm$ 11  $^\circ$& 38$^\circ\pm$10  $^\circ$& 20$^\circ\pm$ 8 $^\circ$\\
     & 1599      $\pm$325          &246$^\circ\pm$ 12  $^\circ$& 39$^\circ\pm$10  $^\circ$& 26$^\circ\pm$ 8 $^\circ$\\\hline
11000  & 1677      $\pm$326          &253$^\circ\pm$ 11  $^\circ$& 38$^\circ\pm$10  $^\circ$& 21$^\circ\pm$ 8 $^\circ$\\
     & 1624      $\pm$325          &246$^\circ\pm$ 12  $^\circ$& 40$^\circ\pm$10  $^\circ$& 26$^\circ\pm$ 8 $^\circ$\\\hline
12000  & 1676      $\pm$326          &253$^\circ\pm$ 11  $^\circ$& 38$^\circ\pm$10  $^\circ$& 21$^\circ\pm$ 8 $^\circ$\\
     & 1624      $\pm$325          &246$^\circ\pm$ 12  $^\circ$& 40$^\circ\pm$10  $^\circ$& 26$^\circ\pm$ 8 $^\circ$\\\hline
13000  & 1652      $\pm$326          &251$^\circ\pm$ 11  $^\circ$& 38$^\circ\pm$10  $^\circ$& 21$^\circ\pm$ 8 $^\circ$\\
     & 1629      $\pm$325          &245$^\circ\pm$ 12  $^\circ$& 39$^\circ\pm$10  $^\circ$& 26$^\circ\pm$ 8 $^\circ$\\\hline
14000  & 1659      $\pm$327          &251$^\circ\pm$ 11  $^\circ$& 38$^\circ\pm$10  $^\circ$& 22$^\circ\pm$ 8 $^\circ$\\
     & 1636      $\pm$326          &245$^\circ\pm$ 11  $^\circ$& 39$^\circ\pm$10  $^\circ$& 26$^\circ\pm$ 8 $^\circ$\\\hline
15000  & 1640      $\pm$327          &251$^\circ\pm$ 11  $^\circ$& 37$^\circ\pm$10  $^\circ$& 21$^\circ\pm$ 8 $^\circ$\\
     & 1633      $\pm$326          &246$^\circ\pm$ 11  $^\circ$& 39$^\circ\pm$10  $^\circ$& 25$^\circ\pm$ 8 $^\circ$\\\hline
16000  & 1638      $\pm$327          &251$^\circ\pm$ 11  $^\circ$& 37$^\circ\pm$10  $^\circ$& 21$^\circ\pm$ 8 $^\circ$\\
     & 1643      $\pm$326          &247$^\circ\pm$ 11  $^\circ$& 38$^\circ\pm$10  $^\circ$& 25$^\circ\pm$ 8 $^\circ$\\\hline
17000  & 1630      $\pm$327          &251$^\circ\pm$ 11  $^\circ$& 37$^\circ\pm$10  $^\circ$& 21$^\circ\pm$ 8 $^\circ$\\
     & 1643      $\pm$326          &247$^\circ\pm$ 11  $^\circ$& 38$^\circ\pm$10  $^\circ$& 25$^\circ\pm$ 8 $^\circ$\\\hline
18000  & 1604      $\pm$328          &251$^\circ\pm$ 11  $^\circ$& 37$^\circ\pm$10  $^\circ$& 21$^\circ\pm$ 8 $^\circ$\\
     & 1631      $\pm$327          &247$^\circ\pm$ 11  $^\circ$& 38$^\circ\pm$10  $^\circ$& 25$^\circ\pm$ 8 $^\circ$\\\hline
19000  & 1591      $\pm$328          &251$^\circ\pm$ 11  $^\circ$& 37$^\circ\pm$10  $^\circ$& 21$^\circ\pm$ 8 $^\circ$\\
     & 1629      $\pm$327          &247$^\circ\pm$ 11  $^\circ$& 38$^\circ\pm$10  $^\circ$& 24$^\circ\pm$ 8 $^\circ$\\\hline
20000  & 1577      $\pm$328          &251$^\circ\pm$ 12  $^\circ$& 37$^\circ\pm$10  $^\circ$& 21$^\circ\pm$ 8 $^\circ$\\
     & 1620      $\pm$327          &247$^\circ\pm$ 11  $^\circ$& 37$^\circ\pm$10  $^\circ$& 24$^\circ\pm$ 8 $^\circ$\\\hline
\label{tab:tab2}
\end{tabular}
\end{center}
\end{table*}

\begin{table*}
\caption[Same as Table 2 but excluding some galaxies] {Same as Table~\ref{tab:tab2} but the analysis excludes the galaxies: Maffei 1, Maffei 2, Dwingeloo 1, IC342 and M81.}
\begin{center}
\begin{tabular}{@{}ccccc}
\hline \\ Dist & $V_{tot}b_L/f(\Omega_{\rm m})$ & l & b &
$\delta\theta$ \\ km $s^{-1}$ & km $s^{-1}$ & deg & deg & deg \\ \hline \\
 1000  &  585      $\pm$254          &280$^\circ\pm$ 30  $^\circ$& 34$^\circ\pm$17  $^\circ$& 25$^\circ\pm$ 9 $^\circ$\\
     &  181      $\pm$228          &360$^\circ\pm$108  $^\circ$& 24$^\circ\pm$29  $^\circ$& 67$^\circ\pm$37 $^\circ$\\\hline
 2000  & 1258      $\pm$282          &263$^\circ\pm$ 13  $^\circ$& 38$^\circ\pm$11  $^\circ$& 15$^\circ\pm$ 9 $^\circ$\\
     &  963      $\pm$261          &259$^\circ\pm$ 18  $^\circ$& 39$^\circ\pm$13  $^\circ$& 19$^\circ\pm$12 $^\circ$\\\hline
 3000  & 1661      $\pm$285          &269$^\circ\pm$  9  $^\circ$& 37$^\circ\pm$ 9  $^\circ$& 11$^\circ\pm$ 5 $^\circ$\\
     & 1352      $\pm$266          &265$^\circ\pm$ 12  $^\circ$& 41$^\circ\pm$10  $^\circ$& 16$^\circ\pm$ 8 $^\circ$\\\hline
 4000  & 1824      $\pm$287          &272$^\circ\pm$  8  $^\circ$& 36$^\circ\pm$ 8  $^\circ$& 10$^\circ\pm$ 4 $^\circ$\\
     & 1571      $\pm$268          &269$^\circ\pm$  9  $^\circ$& 37$^\circ\pm$ 9  $^\circ$& 12$^\circ\pm$ 5 $^\circ$\\\hline
 5000  & 1891      $\pm$288          &272$^\circ\pm$  8  $^\circ$& 33$^\circ\pm$ 7  $^\circ$&  8$^\circ\pm$ 3 $^\circ$\\
     & 1559      $\pm$270          &268$^\circ\pm$  9  $^\circ$& 35$^\circ\pm$ 8  $^\circ$& 10$^\circ\pm$ 5 $^\circ$\\\hline
 6000  & 1678      $\pm$289          &268$^\circ\pm$  8  $^\circ$& 31$^\circ\pm$ 8  $^\circ$&  9$^\circ\pm$ 4 $^\circ$\\
     & 1504      $\pm$271          &266$^\circ\pm$  9  $^\circ$& 32$^\circ\pm$ 8  $^\circ$& 10$^\circ\pm$ 5 $^\circ$\\\hline
 7000  & 1713      $\pm$289          &264$^\circ\pm$  8  $^\circ$& 33$^\circ\pm$ 8  $^\circ$& 11$^\circ\pm$ 6 $^\circ$\\
     & 1551      $\pm$272          &263$^\circ\pm$  9  $^\circ$& 33$^\circ\pm$ 8  $^\circ$& 12$^\circ\pm$ 6 $^\circ$\\\hline
 8000  & 1721      $\pm$290          &264$^\circ\pm$  9  $^\circ$& 35$^\circ\pm$ 8  $^\circ$& 11$^\circ\pm$ 7 $^\circ$\\
     & 1611      $\pm$272          &264$^\circ\pm$  9  $^\circ$& 36$^\circ\pm$ 8  $^\circ$& 12$^\circ\pm$ 7 $^\circ$\\\hline
 9000  & 1704      $\pm$290          &264$^\circ\pm$  9  $^\circ$& 35$^\circ\pm$ 8  $^\circ$& 12$^\circ\pm$ 7 $^\circ$\\
     & 1614      $\pm$272          &264$^\circ\pm$  9  $^\circ$& 36$^\circ\pm$ 8  $^\circ$& 12$^\circ\pm$ 7 $^\circ$\\\hline
10000  & 1691      $\pm$290          &262$^\circ\pm$  9  $^\circ$& 36$^\circ\pm$ 8  $^\circ$& 13$^\circ\pm$ 7 $^\circ$\\
     & 1634      $\pm$272          &262$^\circ\pm$  9  $^\circ$& 36$^\circ\pm$ 8  $^\circ$& 13$^\circ\pm$ 7 $^\circ$\\\hline
11000  & 1691      $\pm$290          &262$^\circ\pm$  9  $^\circ$& 36$^\circ\pm$ 8  $^\circ$& 13$^\circ\pm$ 7 $^\circ$\\
     & 1658      $\pm$273          &262$^\circ\pm$  9  $^\circ$& 37$^\circ\pm$ 8  $^\circ$& 14$^\circ\pm$ 7 $^\circ$\\\hline
12000  & 1690      $\pm$291          &262$^\circ\pm$  9  $^\circ$& 35$^\circ\pm$ 8  $^\circ$& 13$^\circ\pm$ 7 $^\circ$\\
     & 1658      $\pm$273          &261$^\circ\pm$  9  $^\circ$& 37$^\circ\pm$ 8  $^\circ$& 14$^\circ\pm$ 7 $^\circ$\\\hline
13000  & 1665      $\pm$291          &261$^\circ\pm$  9  $^\circ$& 35$^\circ\pm$ 9  $^\circ$& 14$^\circ\pm$ 7 $^\circ$\\
     & 1663      $\pm$273          &261$^\circ\pm$  9  $^\circ$& 36$^\circ\pm$ 8  $^\circ$& 14$^\circ\pm$ 7 $^\circ$\\\hline
14000  & 1671      $\pm$291          &260$^\circ\pm$  9  $^\circ$& 35$^\circ\pm$ 8  $^\circ$& 14$^\circ\pm$ 7 $^\circ$\\
     & 1668      $\pm$273          &260$^\circ\pm$  9  $^\circ$& 36$^\circ\pm$ 8  $^\circ$& 14$^\circ\pm$ 6 $^\circ$\\\hline
15000  & 1652      $\pm$291          &260$^\circ\pm$  9  $^\circ$& 35$^\circ\pm$ 9  $^\circ$& 14$^\circ\pm$ 7 $^\circ$\\
     & 1670      $\pm$274          &261$^\circ\pm$  9  $^\circ$& 36$^\circ\pm$ 8  $^\circ$& 13$^\circ\pm$ 6 $^\circ$\\\hline
16000  & 1653      $\pm$292          &261$^\circ\pm$  9  $^\circ$& 34$^\circ\pm$ 9  $^\circ$& 13$^\circ\pm$ 7 $^\circ$\\
     & 1683      $\pm$274          &261$^\circ\pm$  8  $^\circ$& 35$^\circ\pm$ 8  $^\circ$& 13$^\circ\pm$ 6 $^\circ$\\\hline
17000  & 1647      $\pm$292          &261$^\circ\pm$  9  $^\circ$& 34$^\circ\pm$ 8  $^\circ$& 13$^\circ\pm$ 7 $^\circ$\\
     & 1684      $\pm$274          &261$^\circ\pm$  8  $^\circ$& 35$^\circ\pm$ 8  $^\circ$& 13$^\circ\pm$ 6 $^\circ$\\\hline
18000  & 1619      $\pm$292          &260$^\circ\pm$  9  $^\circ$& 34$^\circ\pm$ 9  $^\circ$& 14$^\circ\pm$ 7 $^\circ$\\
     & 1672      $\pm$274          &261$^\circ\pm$  9  $^\circ$& 35$^\circ\pm$ 8  $^\circ$& 13$^\circ\pm$ 6 $^\circ$\\\hline
19000  & 1606      $\pm$292          &260$^\circ\pm$  9  $^\circ$& 34$^\circ\pm$ 9  $^\circ$& 14$^\circ\pm$ 7 $^\circ$\\
     & 1672      $\pm$274          &262$^\circ\pm$  8  $^\circ$& 35$^\circ\pm$ 8  $^\circ$& 13$^\circ\pm$ 6 $^\circ$\\\hline
20000  & 1594      $\pm$293          &261$^\circ\pm$  9  $^\circ$& 34$^\circ\pm$ 9  $^\circ$& 13$^\circ\pm$ 7 $^\circ$\\
    & 1665      $\pm$275          &262$^\circ\pm$  8  $^\circ$& 34$^\circ\pm$ 8  $^\circ$& 12$^\circ\pm$ 6 $^\circ$\\\hline
\label{tab:tab3}
\end{tabular}
\end{center}
\end{table*}

\section{Discussion}
In this paper, we calculate the 2MRS dipole using number and f\mbox{}l\mbox{}ux
weighting schemes. The f\mbox{}l\mbox{}ux weighted dipole bypasses the effects of redshift space distortions 
and the choice of reference frames giving very robust results.

Our dipole estimates are dominated by the tug of war between the Great Attractor and the Perseus-Pisces superclusters and 
seemingly converge by 6000 \kmps. The contribution 
from structure beyond these distances is negligible. 
The direction of the flux dipole (l=251$^\circ\pm$12$^\circ$,b=37$^\circ\pm$ 10  $^\circ$) is in good agreement with the 2MASS dipole derived by Maller \etal 
(2003) (l=264.5$^\circ\pm$2$^\circ$,b=43.5$^\circ\pm$ 4 $^\circ$). 
The difference in results is probably due to the fact that they use 
a higher latitude cutoff in the mask ($|b|<7^\circ$) 
and exclude all galaxies below this latitude. We confirm this by changing our 
treatment of the Zone of Avoidance to match theirs. We find that 
the flux dipole is very close to their dipole direction. 
Their
limiting Kron magnitude is $K_s=13.57$ which corresponds to an
effective depth of 200 $h^{-1}$ Mpc. As their sample is deep enough to pick
out galaxies in the Shapley Supercluster, the comparison of their dipole
value with our values suggests that the contribution to 
the LG dipole from structure further away 
than the maximum distance of our analysis is not significant. Following Maller \etal (2003), when we adopt $\Omega_{\rm m}=0.27$ we get 
$b_L=1.14\pm0.25$, in good agreement with their value of $b=1.06\pm0.17$ 
suggesting that the 2MRS galaxies are unbiased.
We note that the 2MRS value for the linear bias is somewhat lower than expected
considering that the 2MRS has a high proportion of early type galaxies
which are known to reside mostly in high density regions (e.g. Norberg
\etal 2001, Zehavi \etal 2002). 
The values we derive for $\beta$ and $\omegam$ are consistent with the
concordance $\Lambda$-CDM model values given their error bars. 

Figure 3 shows that the 2MRS
samples the Great Attractor region better than the PSCz survey but
that the PSCz redshift distribution has a longer redshift tail than
the 2MRS. Nevertheless, PSCz dipole agrees 
well with that of the 2MRS. Rowan-Robinson \etal (2000) 
derive a value for $\beta=0.75^{+0.11}_{-0.08}$ 
which is higher than the value we derive in this paper. The PSCz sample is 
biased towards star-forming galaxies and thus under-samples the ellipticals 
which lay in high density regions. Contrarily, the 2MRS is biased towards early-type galaxies. 
This difference may be the reason why they get a higher value 
for $\beta$. The flux weighted dipole is in excellent agreement with the $IRAS$ 1.2 Jy dipole 
(Webster, Lahav \& Fisher 1997) 
which was obtained using a number weighted scheme 
but with an added filter that mitigates the shot noise and deconvolves redshift distortions. 
Our number weighted dipole differs from their results. 
This is probably due to fact that the 2MRS number weighted dipole is plagued wit redshift distortions. Webster, Lahav \& Fisher (1997) obtain the real-space density field from that in the redshift-space using a Wiener Filter. In a forthcoming paper, we will use the same technique to address this issue. Similarly, the 2MRS number weighted dipole also differs from the PSCz dipole 
which was 
calculated using a flow model for massive clusters. 

The analysis of the $IRAS$ QDOT galaxies combined with
Abell clusters (Plionis, Coles \& Catelan 1993) and the X-Ray cluster only dipole (Kocevski, Mullis \& Ebeling 2004 and Kocevski \& Ebeling 2005) imply a significant contribution to the
LG velocity by the Shapley Supercluster. 
Kocevski \& Ebeling (2005) report a significant contribution to 
the LG dipole (56\%) from distances beyond 60 $h^{-1}$ Mpc. 
The discrepancy between their results and ours 
is possibly due to the fact that the 2MRS 
is a better tracer of the galaxies at nearby distances, whereas the
X-ray cluster data are better samplers of the matter distribution beyond 150 $h^{-1}$ Mpc. 

The misalignment angle between the LG and the CMB dipole is smallest at
5000 \kmps where it drops to 12$^\circ\pm$7$^\circ$ 
and increases slightly at larger distances presumably due to shot-noise. 
This behaviour is also observed 
in the other dipole analyses (e.g. Webster, Lahav, Fisher 1997 \& 
Rowan-Robinson \etal 2000). This is a strong indication that most of 
the LG velocity is due to the Great Attractor and 
the Perseus-Pisces superclusters. Of course, we still cannot 
rule out a significant contribution from Shapley as we do not sample that far.
However, it may be more important ask 
what the velocity field in the Great Attractor region is. 
In other words, 
whether we observe a significant backside infall towards the Great Attractor. 

The smallest misalignment angle at 
13000 \kmps is 21$^\circ\pm$8$^\circ$, found for the LG frame, 
f\mbox{}l\mbox{}ux weighted scheme using the second mask. This misalignment can be due to
several effects:
\begin{itemize}
\item The analysis uses linear perturbation theory which is correct
only to first order $\delta$.  There may be contributions to the LG
dipole from small scales which would cause gravity and the velocity
vectors to misalign, even with perfect sampling. However, Ciecielag, 
Chodorowski \& Kudlicki (2001) show that these non-linear effects 
cause only small misalignments for the $IRAS$ PSCz survey. 
However, removing the five most luminous nearby galaxies moves the  
flux dipole to ($l=261^\circ\pm9^\circ$,
$b=34^\circ\pm9^\circ$,cz=20000 \kmps), 8$^\circ$ closer to that of the CMB.  
This suggests that the non-linear effects might be very important in dipole determinations. 
\item The sampling is not perfect and the selection effects of the
surve will increase the shot noise-errors especially at large
distances causing misalignments.
\item There may be uncertainties in the assumptions in galaxy
formation and clustering. For example the mass-to-light ratios might
differ according to type and/or vary with luminosity or the galaxy
biasing might be non-linear and/or scale dependent.
\item There may be a significant contribution to the LG dipole from
structure further away than the maximum distance of our analysis.
\item The direction of the LG dipole may be affected by nearby galaxies
at low latitudes which are not sampled by the 2MRS. In the future, the
masked regions will be filled by galaxies from other surveys such as
the ongoing HI Parkes Deep Zone of Avoidance Survey (Henning \etal
2004) as well as the galaxies that are sampled from the 2MRS itself.
\end{itemize}
Our initial
calculations of the expected LG acceleration (c.f. Lahav, Kaiser \&
Hoffman 1990; Juszkiewicz, Vittorio \& Wyse 1990) suggest that the
misalignment of 21$^\circ\pm$8$^\circ$ is within 1 $\sigma$ of the
dipole probability distribution in a CDM Universe with $\Omega_{\rm
m}=0.3$.
In a forthcoming paper, the cosmological parameters will be
constrained more vigourously using a maximum likelihood analysis based
on the spherical harmonics expansion (e.g. Fisher, Scharf \& Lahav
1994; Heavens \& Taylor 1995) of the 2MRS density field.  

\section*{ACKNOWLEDGEMENTS}
We thank Sarah Bridle, Alan Heavens, Ariyeh Maller and Karen Masters for their
useful comments.  
PE would like thank the University College London for its hospitality
during the completion of this work. OL acknowledges a PPARC Senior
Research Fellowship.  JPH, LM, CSK, NM, and TJ are supported by NSF
grant AST-0406906, and EF's research is partially supported by the Smithsonian Institution.  
DHJ is supported as a Research Associate by
Australian Research Council Discovery-Projects Grant (DP-0208876),
administered by the Australian National University.  This publication
makes use of data products from the Two Micron All Sky Survey, which
is a joint project of the University of Massachusetts and the Infrared
Processing and Analysis Center/California Institute of Technology,
funded by the National Aeronautics and Space Administration and the
National Science Foundation. This research has also made use of the NASA/IPAC Extragalactic Database (NED) 
which is operated by the Jet Propulsion Laboratory, California Institute of Technology, under contract 
with the National Aeronautics and Space Administration and the SIMBAD database,
operated at CDS, Strasbourg, France.

{}

\end{document}